\documentclass[conference]{IEEEtran}

\usepackage[utf8]{inputenc}

\usepackage{cite}
\usepackage{textcomp}
\usepackage{xcolor}

\usepackage{amsfonts}
\usepackage{amsmath}
\usepackage{algorithm} %not IEEE friendly
\usepackage{algpseudocode}
\usepackage{graphicx}
\usepackage{amssymb}
\usepackage{amsthm}
\usepackage{mathtools}
\usepackage{bm}
\usepackage{hyperref}
\usepackage{adjustbox}   
\usepackage{subfig}
\usepackage{tabularx,colortbl}
\usepackage{threeparttable}
\usepackage{multirow}
\usepackage[normalem]{ulem}
\usepackage [english]{babel}
\newcommand{\TT}{\begin{tt}tensor-tools\end{tt}}

\newcommand{\vcr}[1]{\ensuremath{\bm{#1}}}                   % vector
\newcommand{\mat}[1]{\ensuremath{\bm{#1}}}                     % matrix
\newcommand{\tsr}[1]{\ensuremath{\bm{\mathcal{#1}}}}                     % tensor
\newcommand{\vel}[1]{\lowercase{#1}}                     % vector element
\newcommand{\mel}[1]{\lowercase{#1}}                     % matrix element
\newcommand{\tel}[1]{\lowercase{#1}}                  % tensor element

\usepackage{tikz}
\usetikzlibrary{calc}
\begin{document}

\title{Distributed-Memory DMRG via Sparse and Dense Parallel Tensor Contractions}

\author{\IEEEauthorblockN{Ryan Levy\IEEEauthorrefmark{1}\IEEEauthorrefmark{2},
Edgar Solomonik\IEEEauthorrefmark{1}\IEEEauthorrefmark{3}, and Bryan K. Clark\IEEEauthorrefmark{1}\IEEEauthorrefmark{2}
}
\IEEEauthorblockA{\IEEEauthorrefmark{1}\textit{Illinois Quantum Information Science and Technology Center (IQUIST)} \\ \IEEEauthorrefmark{2}\textit{Institute for Condensed Matter Theory and Department of Physics} \\ \IEEEauthorrefmark{3}\textit{Department of Computer Science}\\
\textit{University of Illinois at Urbana-Champaign, IL 61801 USA} \\
Email: \{rlevy3, solomon2, bkclark\}@illinois.edu}}

\maketitle

\begin{abstract}
The density matrix renormalization group (DMRG) algorithm is a powerful tool for solving eigenvalue problems to model quantum systems.
DMRG relies on tensor contractions and dense linear algebra to compute properties of condensed matter physics systems.
However, its efficient parallel implementation is challenging due to limited concurrency, large memory footprint, and tensor sparsity.
We mitigate these problems by implementing two new parallel approaches that handle block sparsity arising in DMRG, via Cyclops, a distributed memory tensor contraction library. 
We benchmark their performance on two physical systems using the Blue Waters and Stampede2 supercomputers. 
Our DMRG performance is improved by up to 5.9X in runtime and 99X in processing rate over ITensor, at roughly comparable computational resource use.
This enables higher accuracy calculations via larger tensors for quantum state approximation.
We demonstrate that despite having limited concurrency, DMRG is weakly scalable with the use of efficient parallel tensor contraction mechanisms.

\end{abstract}


\section{Introduction} 

One of the most successful optimization algorithms for 1D systems, the density matrix renormalization group (DMRG) \cite{White1992,White1993,Schollwock2011}, is celebrated for its speed and quality. By representing the Hamiltonian matrix as a series of tensor products, known as a matrix product operator (MPO), the ground state, i.e. minimal eigenvector of the Hamiltonian, can also be efficiently represented by tensor products in 1D. 
When applied to 2D systems, one must linearize the system, resulting in harder problems that scale exponentially as the width of the system.  Large scale DMRG is then needed to effectively solve these problems. 
Achieving higher accuracy in DMRG enables better characterization of properties of fundamental quantum physical models of strongly correlated materials.
However, high accuracy requires working with very large sparse tensors, necessitating the use of supercomputing resources.

The cost associated with accurate simulations of 2D systems have spurned efforts to improve parallelism within DMRG\cite{Stoudenmire2013,ueda2018infinite};  so far the most practical improvements have been in the area of shared memory parallelism\cite{dolfi2014matrix,hager2004parallelization} and the plurality of 2D DMRG papers use this mode of parallelism (if any parallelism at all).   Two broad attempts at making a distributed-memory parallel DMRG have involved parallelizing the tensors themselves\cite{chan2004algorithm,Yamada2009, Vance2017,Dolfi2019,Kantian2019,Brabec2020} or developing a new numerical formulations of DMRG that allow for more concurrency by trading-off accuracy and compromising monotonicity of optimization.
We demonstrate that effective use of distributed tensor contraction primitives suffices to accelerate the traditional DMRG algorithms with HPC resources and enable cost-effective high-accuracy calculations.

These improvements are consequential for the state of practice. A large-scale DMRG simulation can often take many weeks on a single node and is limited in accuracy by the available RAM on a machine.
A massively parallel code which uses a distributed memory paradigm can overcome both these obstacles, accelerating the wall-clock time to achieve scientific results from weeks to days and reaching previously inaccessible wave function quality.
However, parallelization is complicated by the need to exploit block-sparsity in tensors (due to symmetries in different systems) to minimize memory footprint and computational cost.

We present an implementation of the DMRG algorithm using the Cyclops Tensor Framework\cite{solomonik2014ctf}, allowing for a massively parallel code that has been observed to reach up to 3 TFlops/s on state of the art problems. We focus on finite 2D lattice models, which traditionally have an easily constructed Hamiltonian but require significant computational time to converge. 

We introduce the formalism of tensor networks and provide a description of the DMRG algorithm in Section~\ref{sec:bgnd}.
Section~\ref{sec:bgnd} describes prior work and tabulates past studies of parallel DMRG.
Our own work is the first comparative study of DMRG parallelization approaches.
In particular, In Section~\ref{sec:alg}, we provide three approaches for managing block sparsity via sparse and dense distributed tensors.
In Section~\ref{sec:systems}, we introduce two model problems characterizing different workloads: the 2D $J_1-J_2$ Heisenberg model at $J_2=0.5$ (abbreviated \textit{spins} throughout this work) and the triangular Hubbard model (abbreviated \textit{electrons}). Our numerical experiments in Section~\ref{sec:exp} demonstrates speed-ups of up to 99x in performance rate relative to a state-of-the-art single node code, with roughly the same resource efficiency.

\section{DMRG Background}\label{sec:bgnd}

We provide an overview of the DMRG algorithm and quantum number symmetries (specifically of $U(1)$ symmetries), which describe the sparsity structure of tensors in DMRG.
More comprehensive reviews are available for tensor networks~\cite{Orus2014}, DMRG~ \cite{White1992,White1993,Schollwock2011} and quantum number symmetries~\cite{Orus2014b}.
At a high-level, given a Hermitian matrix represented as a 1D tensor network, the DMRG algorithm seeks to compute the eigenpair with the smallest eigenvalue (ground state energy) by using alternating optimization of a 1D tensor network that approximately represents the eigenvector.

\subsection{Tensors, Tensor Networks, and Tensor Diagrams}
\begin{figure}[t]
\centering %trim=left bottom right top
\includegraphics[trim=4cm 0.6cm 2cm 0cm,clip,width=\columnwidth,]{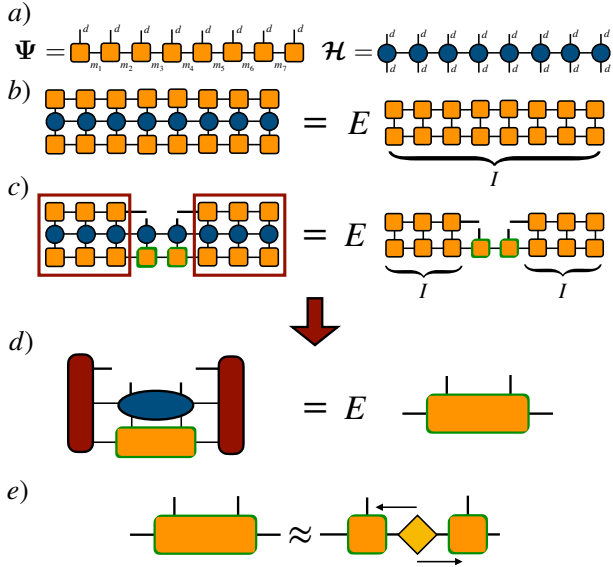}%trim=5cm 1cm 4cm 2cm,clip,

\caption{$a)$ \textit{left} the matrix product state (MPS) $\bm \Psi$ in orange as a tensor network and \textit{right} the Hamiltonian $H$ as a matrix product operator (MPO) in blue as a tensor network $\tsr{H}$. Annotated are the physical indices $d$ and the bond dimensions of the MPS $m_i$. 
$b)$  The original problem $\langle \bm{\Psi}| \tsr H |\bm{\Psi}\rangle = E \langle\bm{\Psi}|\bm{\Psi}\rangle $ with constraint shown in the tensor network representation. 
$c)$ We select two adjacent sites to optimize simultaneously, shown in green. The optimization problem for these sites with normalization constraint is then recast as an eigenvalue problem. By exploiting an extra degree of freedom of the MPS, certain contractions can be reduced to the identity.
$d)$ The full optimization problem in $b$ is never directly used, instead an efficient representation is created by contracting all other sites into left and right environments. The two site tensor, shown in orange, is then optimized via a Davidson routine. 
$e)$ After optimization, the order-4, two site tensor is split using SVD and (potentially) truncated to a bond dimension $m_j$. The singular values can then be absorbed either left or right, following the sweep direction, in order to retain a proper the orthogonal structure. }
\label{fig:dmrg_diag}
\end{figure}

The DMRG algorithm works with complex tensors.
We denote an \emph{order} $N$ (with $N$ \emph{modes}) tensor of \emph{dimensions} $s_1\times \cdots \times s_N$ as $\tsr T\in\mathbb{C}^{s_1\times \cdots \times s_N}$ and its elements as $\tel{T}_{i_1\cdots i_N}$.
A \emph{tensor network} $\bm{f}_G(\tsr{T}^{(1)},\cdots,\tsr{T}^{(M)})$ is described by a multi-graph $G=(V,E)$, where $V=\{\tsr{T}^{(1)},\cdots,\tsr{T}^{(M)}\}$ and the edges denote indices that define contraction between a pair of modes of two tensors or an uncontracted index (which we represent by a loop).
If $E_i=\{e_{i1},\cdots, e_{im_i}\}\subseteq E$ is the collection of edges adjacent to vertex $i$ and $L=\{l_1\cdots l_K\} \subseteq E$ is the set of loops, we can write the tensor network function as
\[w_{l_1\cdots l_k} = \sum_{e \in E} \prod_{j=1}^M \tel{T}^{(j)}_{e_{i1} \cdots e_{im_i}}.\]
Any such tensor contraction can be mapped to a matrix multiplication.
If matrix multiplication is performed using the classical $O(n^3)$ algorithm, the cost of the contraction is given by the product of the dimensions of the tensor modes corresponding to all of the indices in $E$.
We leverage the Einstein summation convention, omitting summation indices to describe tensor contractions, for instance we describe matrix multiplication as $\mel{c}_{ij}=\mel a_{ik} \mel b_{kj}$.

A \emph{tensor diagram} is a depiction of a tensor network $\bm{f}_G$ via the graph $G$, except that instead of loops, uncontracted edges correspond to edges that point into whitespace.
Tensor diagrams provide a precise and intuitive way of expressing tensor networks, through which it is easier to see both the geometric structure as well as reason about contraction orderings, than via the algebraic expression of the tensor contraction.
Tensor diagrams are widely used in tensor network literature; we refer the reader to~\cite{Orus2014} for a comprehensive introduction to their applications and interpretation.

\subsection{Matrix Product States}

The DMRG algorithm uses 1D tensor networks, namely the \emph{matrix product state} (MPS) and the \emph{matrix product operator} (MPO)~\cite{Schollwock2011}.
These tensor networks are referred to as \emph{tensor trains} in literature on tensor decompositions~\cite{oseledets2011tensor}.
The MPS and MPO are used to represent a vector (the eigenvector guess in DMRG) and a matrix (the Hamiltonian in DMRG), respectively.
Figure~\ref{fig:dmrg_diag}a provides the tensor diagram for an MPS (\textit{left}) and  displays the tensor diagram for an MPO (\textit{right}).

The MPS is used as an approximation of the sought-after eigenvector in DMRG, also referred to as the \emph{wave function}.
The MPS tensor network contracts into an order $N$ tensor $\bm{\Psi}\in \mathbb{C}^{d\times \cdots\times d}$ described by a set of $N$ \emph{sites}, each of which is represented by an order three tensor\footnote{the first and last tensor would have one mode be of unit dimension by this convention} $\tsr{T}^{(j)}$ whose elements are $\tel{T}^{(j)}_{i_j \sigma_j i_{j+1}}$.
The MPS contracts to yield a tensor that corresponds to a \emph{folding} of a vector $\text{vec}(\bm{\Psi})\in\mathbb{C}^{d^N}$,
\[
\tel{\psi}_{\sigma_1\cdots \sigma_N} = \sum_{i_1\cdots i_{N+1}} \prod_{j=1}^N \tel T^{(j)}_{i_j \sigma_j i_{j+1}}.
\]
The MPO represents the Hamiltonian matrix $\mat{H}$ as a tensor $\tsr H$ of order $2N$ and factorizes it into a product of order 4 tensors,
\[
\tel{H}_{\sigma_1\cdots \sigma_N \nu_1\cdots \nu_N } = \sum_{i_1\cdots i_{N+1}} \prod_{j=1}^N \tel H^{(j)}_{i_j \sigma_j\nu_j i_{j+1}}.
\]
The \emph{bond dimension} of an MPS or an MPO, which we denote by $m$ and $k$ respectively, is the maximum dimension of a mode of any site indexed by a contracted index $i_j$.
We also refer to the $j$th bond dimension (range of index $i_j$) of an MPS as $m_j$.
The \emph{physical dimension} of an MPS or an MPO, which we denote by $d$, is the dimension of each mode of $\bm{\Psi}$ and $\tsr{H}$ and in many contexts is a fixed small number like $2$ or $4$.
The product of an MPO and an MPS $\tsr H|\bm{\Psi}\rangle$ can be represented exactly as an MPS with bond dimension $kd$.

MPOs with low bond dimension provide an exact representation of Hamiltonians for quantum lattice systems with local interactions~\cite{Schollwock2011}.
The bond dimension generally grows with the number of terms in the Hamiltonian, which depends on the physical system being studied.

\subsection{DMRG Algorithm}

Given an MPO representation of a Hamiltonian $\mat H$, the DMRG algorithm seeks to approximate its ground state and energy thereof by 
minimizing
\[\min_{\bm \Psi\in \mathbb{C}^{d\times \cdots \times d}, \langle \bm \Psi, \bm \Psi \rangle = 1} \underbrace{\langle \bm \Psi | \tsr H | \bm \Psi \rangle}_{\text{vec}(\bm \Psi)^\dagger \mat H\text{vec}(\bm \Psi)}.\]
with $\bm \Psi$ approximated by an MPS. 
To do so the DMRG algorithm optimizes each site in an alternating manner.
In particular it updates each site $\tsr{T}^{(j)}$ of the MPS by solving a reduced quadratic optimization problem,
\[\min_{\tsr{T}^{(j)}\in\mathbb{C}^{m_j\times d\times m_{j+1}},\langle \tsr{T}^{(j)},  \tsr{T}^{(j)} \rangle = 1}\text{vec}(\tsr{T}^{(j)})^{\dagger} \underbrace{\mat{Q}{}^\dagger\mat{H} \mat{Q}}_{\mat K} \text{vec}(\bm{\tsr{T}^{(j)}}).\]
The projection $\mat{Q}$ is a matricization of a tensor  $\tsr{Q}$ defined by the contraction of all sites in the MPS except $\tsr{T}^{(j)}$,
\[
\tel{Q}_{\sigma_1\cdots \sigma_N i_{j}i_{j+1} } = \sum_{i_1\cdots i_{j-1}i_{j+2}\cdots i_{N+1}} \prod_{k=1,k\neq j}^N \tel T^{(k)}_{i_k \sigma_k i_{k+1}}.
\]
The projected Hamiltonian $\mat{K}\in\mathbb{C}^{m_jdm_{j+1}\times m_jdm_{j+1}}$ is the matricization of a tensor $\tsr{K}\in\mathbb{C}^{m_j\times d \times m_{j+1} \times m_j \times d \times m_{j+1}}$, so that
\[\tel{K}_{i_j\sigma_j i_{j+1} l_j \nu_j l_{j+1}}=\sum_{\substack{\sigma_1\cdots \sigma_{j-1}\sigma_{j+1}\cdots \sigma_n \\ \nu_1\cdots \nu_{j-1}\nu_{j+1}\cdots \nu_n}}\tel{Q}{}^{\dagger}_{\bm{\sigma} i_{j}i_{j+1}}\tel{H}_{\bm{\sigma}\bm{\nu}} \tel{Q}_{\bm{\nu} l_{j}l_{j+1}},\]
where we use vector notation for indices, e.g., $\bm{\sigma} = \sigma_1\cdots\sigma_N$.
In matrix form, we can write $\tsr{Q}$ as \(\mat{Q}=\mat{L} \otimes \mat{I} \otimes \mat{R}\).
The left and right components of the MPS $\mat{L}$ and $\mat{R}$ may be orthogonalized $\mat{L}$ by performing a QR factorization of each site and maintaining orthogonality during the optimization process, which gives a QR factorization of $\mat{L}$ overall, after which the upper-triangular factor may be absorbed into the tensor $\tsr{T}^{(j)}$.
When $\mat{L}$ and $\mat{R}$ are both orthogonal, the MPS is said to be in a \emph{canonical form} with \emph{center site} $j$.
A canonical form ensures that $\mat{Q}$ is an orthogonal projection and that any approximation (local error) to $\tsr{T}^{(j)}$ amplifies overall error in the state minimally~\cite{zhang2020stability}.
These components of the MPS are contracted with respective parts of the MPO $\tsr{U}$ and $\tsr{W}$, which include all sites in the MPO before and after the $j$th site.
Within DMRG, these are combined with $\tsr{L}$ and $\tsr{R}$, respectively, to form the \emph{left} and \emph{right environments}.
For example, the left environment is given by
\begin{align*}
\tel{A}_{i_jk_jl_j} =& \sum_{\sigma_1\cdots \sigma_{j-1}}\sum_{\nu_1\cdots \nu_{j-1}}\tel{L}{}^{\dagger}_{\sigma_1\cdots \sigma_{j-1}i_j}\\
&\underbrace{\bigg(\sum_{k_1\cdots k_{j-1}}\prod_{n=1}^{j-1} \tel{H}^{(n)}_{k_n\sigma_n\nu_nk_{n+1}}\bigg)}_{\tel{U}_{\sigma_1\cdots \sigma_{j-1}\nu_1\cdots \nu_{j-1}k_{j}}}\tel{L}_{\nu_1\cdots \nu_{j-1}l_j}.
\end{align*}
The right environment, $\tsr{B}$ is formed similarly, and then the reduced Hamiltonian is defined by
\[\tel{K}_{i_j\sigma_j i_{j+1} l_j \nu_j l_{j+1}} = \sum_{k_{j},k_{j+1}}\tel{A}_{i_jk_jl_j}\tel{H}^{(j)}_{k_j\sigma_j\nu_jk_{j+1}}\tel{B}_{i_{j+1}k_{j+1}l_{j+1}}.\]
This representation of $\tsr{K}$ 
is dominated in size by $\tsr{A}$ and $\tsr{B}$ both of which have $m^2k$ elements, while $\tsr{K}$ is of size $m^4d^2$.
DMRG sweeps left-to-right and then back, extending the environments from center to center by contracting with the updated tensor.
After forming the environments, the updated $\tsr{T}^{(j)}$ may be obtained using an iterative method such as CG or Davidson's algorithm with $\mat{K}$ applied via the implicit form defined by the two environments and center MPO site $\tsr{H}^{(j)}$.
The form of each matrix vector product, which has overall cost $O(m^3kd)$, is described by the tensor diagram in fig.~\ref{fig:dmrg_diag}(d).
\begin{table*}[t]
    \centering
    \resizebox{\textwidth}{!}{ %make table fit without guessing
    \begin{threeparttable}[t]
\begin{tabular}{llllrr}
 &  &  &  & Maximum & Maximum\tabularnewline
System & Work & Method & Architecture & Bond Dim. ($m$) & Nodes\tabularnewline
\hline 
\hline 
 &  &  &  &  & \tabularnewline
Heisenberg $J_{1}-J_{2}$ & \textbf{this work} & $U(1)$ DMRG & Distributed Memory & $32\thinspace768$ & 256\tabularnewline
 & Jiang, et al.\cite{Jiang2012} &  & NR\tnote{1} & $12\thinspace000$ & NR\tnote{1}\tabularnewline
 & Wang, et al.\cite{Wang2018} &  &  & $12\thinspace000$ & \tabularnewline
 &  &  &  &  & \tabularnewline
Triangular Hubbard & \textbf{this work} & $U(1)$ DMRG & Distributed Memory & $32\thinspace768$ & 256\tabularnewline
 & Shirakawa, et al.\cite{Shirakawa2017} &  & NR\tnote{1} & $20\thinspace000$ & NR\tnote{1}\tabularnewline
 & Szasz, et al.\cite{Szasz2018} & $U(1)+k$ space iDMRG & Shared Memory & $11\thinspace314$ & \tabularnewline
 &  &  &  &  & \tabularnewline
Hubbard 1D Chain & Rinc{\'{o}}n, et al.\cite{Rincon2010} & \multirow{2}{*}{$U(1)$ DMRG} & Distributed Memory\tnote{2} & $1\thinspace000$ & 8\tabularnewline
$U-V$ Hubbard & Kantian, et al.\cite{Kantian2019,Dolfi2019} &  & Distributed Memory & $18\thinspace000$ & 180\tnote{3}\tabularnewline
Square Hubbard & Yamada, et al.\cite{Yamada2011,Yamada2009} & $s-$leg DMRG & Distributed Shared Memory & $1\thinspace200$ & \tabularnewline
 &  &  &  &  & \tabularnewline
Heisenberg 1D Chain & Vance, et al.\cite{Vance2017} & $U(1)$ iDMRG & Distributed Memory\tnote{4} & $2\thinspace048$ & 64\tabularnewline
Heisenberg $J_{1}$ & Stoudenmire, et al.\cite{Stoudenmire2013} & Parallel $U(1)$ DMRG & Real-Space Parallel & $2\thinspace000$ & 10\tabularnewline
\tabularnewline
\end{tabular}

\begin{tablenotes}
    \item [1] Not Reported, assumed single node shared memory architecture
    \item [2] blocks are distributed but elements are not distributed over processors
    \item [3] we use the timing results published in ref.~\cite{Dolfi2019} which we expect to correlate with the publication ref.~\cite{Kantian2019}
    \item [4] via PETSc and SLEPc to ref.~\cite{Kantian2019}
\end{tablenotes}
\end{threeparttable}
    }
    \caption{Comparison of prior work on the systems of interest and other known parallel DMRG works on the lattice. The physical system are 2D cylinders unless noted otherwise. For completeness, we include two infinite DMRG (iDMRG) methods\cite{White1992,White1993,Stlund1995}, which, while similar in some aspects to DMRG, is a different algorithm beyond the scope of this work.   }
    \label{tab:bond_dim_comp}
\end{table*}

%%%%%%%%%%%%%%%%%

Our implementation of the Davidson algorithm \cite{Davidson1975} is based on the ITensor library implementation \cite{itensor}, except without the use of preconditioning and with randomization to alleviate failed reorthogonalization.
Alg.~\ref{alg:JD} outlines the approach in matrix form.

\begin{algorithm}
    \begin{algorithmic}[1]
    \small
    \State{\textbf{Input: }Tensor $\vcr x_0\in\mathbb{R}^{s_1\times\cdots\times s_4}$}
    \State{Initialize $\vcr v_0=\vcr x_0$, $\vcr v^A_0=\mat A\vcr x_0$, matrix \mat{M}}
    \For{i=0,1,...}
        \For{j=0,\dots,i}
        \State{$m_{ij}=m_{ji}= (\vcr v^{A}_j)^\dagger \vcr v_i$}
        \EndFor
        \State{Diagonalize leading $i\times i$ block of $\mat M$, computing smallest eigenvalue/vector $(\lambda,\vcr s)$} 
        \State{$\vcr x=\sum_j s_j \vcr v_j, \vcr q= \sum_j s_j \vcr v^A_j$}
        \State{$\vcr q=\vcr q-\lambda \vcr x$}
        \State{Check convergence based on norm of $\vcr q$} 
        \State{Orthogonalize $\vcr q$ with all $\vcr v_j$ via modified Gram-Schmidt}
        \State{$\vcr v_{i+1}=\vcr q$,\quad $\vcr v^A_{i+1} = \mat A\vcr v_{i+1}$}
    \EndFor
    
    \State{\Return $\vcr x/\left\|\vcr x\right\|$}
    \end{algorithmic}
    \caption{Davidson routine implementation. The effective matrix $\mat A$ given by the MPO and environment tensors, is shown in fig.~\ref{fig:dmrg_diag} d).}
    \label{alg:JD}
\end{algorithm}

In doing DMRG, we gradually increase bond dimension of the MPS, sweeping over all sites multiple times for each successive bond dimension choice.
During the sweep, we use a subspace size of 2 in the Davidson routine. Additionally, while preconditioning accelerates convergence, we find that for the problems presented here, the additional memory and time cost is prohibitive compared to the cost of running more sweeps. This can be attributed to fact that each optimization subproblem is supplied a very good initial guess and need not be solved to high accuracy for an intermediate bond dimensions.
%%%%%%%%%%%%%%%%%
 
A standard extension of optimizing a single site is to optimize two sites simultaneously. By contracting two neighboring sites and forming 
\begin{align*}
\vel{x}^{(j,j+1)}_{i_j \sigma_j \sigma_{j+1} i_{j+2}} %=& \tsr{T}^{(j)}_{i_j \sigma_j i_{j+1}}\tsr{T}^{(j+1)}_{i_{j+1} \sigma_{j+1} i_{j+2}} \\
=& \sum_{i_{j+1}} \tel{T}^{(j)}_{i_j \sigma_j i_{j+1}} \tel{T}^{(j+1)}_{i_{j+1} \sigma_{j+1} i_{j+2}}
\end{align*}
and then performing optimization over $\vcr{x}^{(j,j+1)}$.  

After optimization, the new $\vcr{x}^{(j,j+1)}$ is decomposed back into two tensors via singular value decomposition (SVD) shown pictorially in fig.~\ref{fig:dmrg_diag}e. The number of singular values kept determines the new bond dimension of index $i_{j+1}$ shared on the two order-3 tensors of the MPS $\tsr{T}^{(j)}$,$\tsr{T}^{(j+1)}$.
The bond dimension at site $j$ can increase exponentially from the two ends, so $m_j\leq \min(d^j,d^{N-j})$, but is generally capped at a particular value, $m\sim O(10^3-10^4)$. Error incurred due to truncation can be calculated from the sum of the truncated singular values. Our implementation removes all singular values below $10^{-12}$.  Note that the MPO bond dimension\footnote{e.g. we consider $k\sim 30$ and $m>4096$} $k \ll m$. 

 \subsection{Quantum Numbers}
 An essential improvement is to decompose the tensors by global symmetry representations. Consider a global symmetry group $G$ with element $g\in G$ and unitary operator representation $U_g \in \mathbb{C}^{d^{2N}}$. We will restrict to abelian groups, namely $U(1)$, although it is possible to consider non-abelian groups as well. 
Most Hamiltonians of interest respect a global symmetry, i.e. $[H,U_g]=0$ $\forall g\in G$. 
Examples of the symmetries include the total magnetic spin $S_z$ and/or particle number.
This permits the Hamiltonian, and subsequently the MPS and MPO, to be represented into a \textit{block form} (see cartoon in fig.~\ref{fig:structure_cartoon}b). 
Once a given global symmetry is fixed in the MPS, the Hamiltonian and all local operators will respect this symmetry by Schur's lemma, and DMRG will respect the block form of the tensor networks.

 Each order-$r$ tensor $\tsr{T}$ can be described in block form with a list $\{q^{(\ell)}\}$ of quantum number label tuples $q^{(\ell)}\in \mathbb{Z}^{r}$ where each tuple of labels correspond to an independent order-$r$ tensor block $T_{q^{(\ell)}} \in \mathbb{C}^{d^\ell_1\times d^\ell_2\times\dots\times d^\ell_r}$. 
 Each label $q^{(\ell)}_i$ in turn corresponds to a different degenerate space, and thus the size of each tensor block index is variable with a maximum size determined by the group symmetry element. 
 With every block index set to the maximum size we can recover the original space of the tensor network.  
 
 Using quantum blocks induces block-sparsity through the tensors; the nature of the block sparsity  (i.e. size, number, and structure of blocks) depends on the physics of the system (as seen in fig.~\ref{fig:block_sizes}).   Quantum blocks improves DMRG in various ways. When represented in a sparse or block-sparse format, the required memory can be decreased from $\prod_i^r d_i$ to $\sum_\ell \prod_i^r d^\ell_i$ per tensor.
 Contractions and SVD can now be performed over individual blocks.  As the cost of these operations is often cubic in bond dimension, this is a non-trivial savings.

\section{Prior Work}

Despite the requirement for large DMRG simulations in various physics problems (the DMRG algorithm is cited $\approx$6000 times), there has been little widespread use of a massive parallel implementation on the lattice.  This situation differs in quantum chemistry applications of DMRG  where there are higher costs associated with individual tensor contractions due to basis sets\cite{chan2004algorithm,Brabec2020}.

In table~\ref{tab:bond_dim_comp}, we compare and contrast the prior work both on other parallel DMRG attempts, as well as serial DMRG simulations on the two prototypical physics systems we consider, the two-dimensional Heisenberg $J_1-J_2$ and the triangular Hubbard system; these systems are further discussed in section~\ref{sec:systems}.

It is interesting to note that the plurality of simulations with the largest bond dimensions to date have primarily been on serial or shared memory machines.  In particular, all studies of the two systems we are considering fall into this category.  Bond dimensions have saturated around $m\sim 10\thinspace 000$ and are quickly being limited by the RAM required to store the necessary tensors and intermediate contractions.  To avoid an extra factor of system size in RAM, the tensors for all but the two sites being worked on are often written to disk; this generates additional significant latency.  At some point, even storing the tensors in RAM for a single optimization becomes impossible in the RAM on a single node.  Our work is the first to exploit parallelism via general sparse and dense distributed tensor contractions. Our approach not only drastically improves the wall-clock time of the calculation over single-node execution, but gives access to bond dimensions inaccessible to the RAM on a single node (avoiding even the need to write to disk). 

There are previous works that have implemented distributed-memory parallel DMRG and DMRG like algorithms. The cases where the standard DMRG algorithm has been parallelized with distributed memory include refs.~ \cite{Vance2017,Rincon2010,Kantian2019} and massively-parallel shared memory in ref.~\cite{Yamada2011}.

\begin{figure}
    \centering
    \subfloat[Block Structure\label{fig:block_sizes}]{\includegraphics[width=0.5\columnwidth]{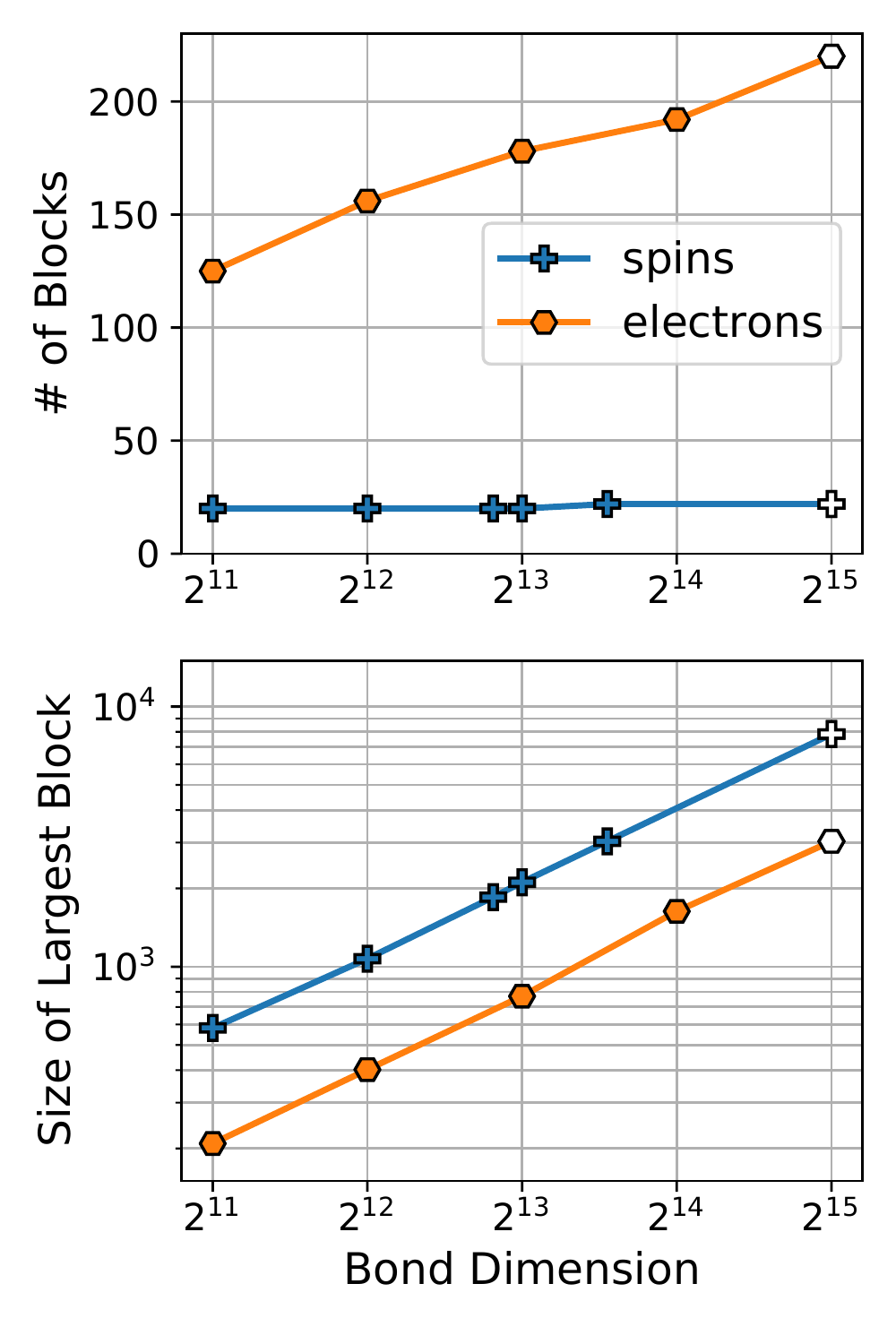}}
    \subfloat[Sparsity Comparison\label{fig:sparsity}]{\includegraphics[width=0.5\columnwidth]{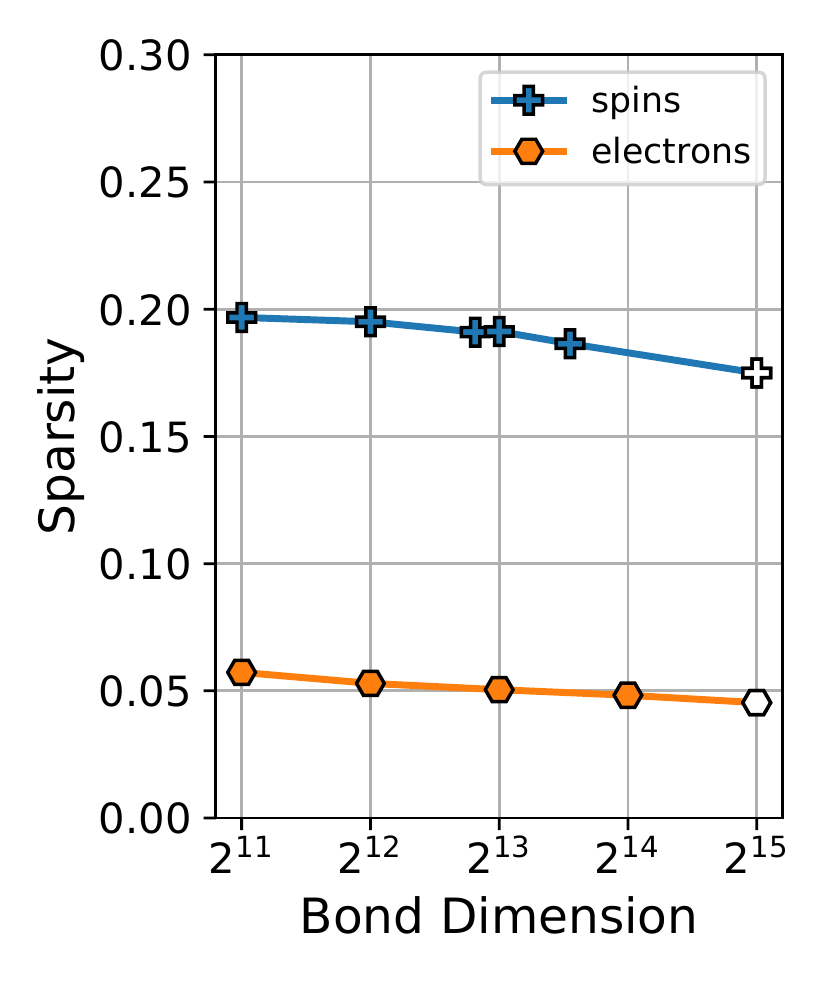}}
    \caption{comparison of MPS  \textit{(a)} block structure and \textit{(b)} sparsity for a representative MPS tensor. Open circle results 
    are taken from MPS where only a small subset of sites (including the measured site) were at a given bond dimension.  The largest block scales as $m^{0.94}$ for spins and $m^{0.97}$ for electrons.}  
\end{figure}

Ref.~\cite{Rincon2010} achieves parallelism over a one-dimensional system by distributing different quantum number blocks to different nodes.  As the size of the largest block  generically scales linearly with the bond dimension (see fig.~\ref{fig:block_sizes}), this severely limits the maximum achievable problem size.   Refs.~\cite{Kantian2019,Dolfi2019} develop a parallel distributed memory implementation to study the $U-V$ Hubbard model ($d=4$).  Using threading with Intel\textsuperscript{\textregistered} Cilk\textsuperscript{\texttrademark} Plus, they scale to a bond dimension of $m= 18\thinspace 000$ via block-sparse matrix contractions parallelizing blocks over processors and block elements over groups of processors.  This approach was outlined in the appendix of a paper \cite{Kantian2019} describing new physics on the UV-model.
Some scaling results of this effort are included in a related thesis \cite{Dolfi2019}.
In ref.~\cite{Yamada2011}, the authors use a slightly-modified DMRG algorithm; instead of optimizing two sites at a time, they optimize  over $2W$ sites simultaneously where $W$ is the width of their problem.  This induces a cost of $W$ over standard DMRG which they deal with in a parallel manner.

Other approaches instead parallelize alternatives or variants to DMRG.  Ref.~\cite{Vance2017} parallelizes infinite DMRG\cite{White1992,White1993,Stlund1995}, a variant of DMRG designed for translationally invariant MPS of uniform tensors $\tsr{T}=\tsr{T}^{(j)}~\forall j$, by parallelizing contractions using the PETSc and SLEPc library.
In ref.~\cite{Stoudenmire2013} a DMRG-like algorithm is proposed in which different nodes work on different ranges of sites.   While this approach is shown to achieve good parallel scalability with over 10 nodes, each optimization is done in a way that is not consistent with the tensors on other nodes, resulting in potential loss of accuracy and monotonicity in optimization.

Our work differs from these previous works on distributed DMRG both in scale as well as approach.  We compute DMRG in the same way as the best sequential approach, preserving both the efficiency (i.e. same number of flops) as well as benefits of the standard algorithm.  Unlike some prior approaches, we directly distribute each tensor (or quantum block of a tensor) over all nodes.  This allows each processor to work simultaneously on each contraction and avoids load-balancing issues associated with different size quantum blocks on different nodes.   In terms of scale, we use significantly more nodes (256), tackle difficult two-dimensional systems, and reach significantly bigger bond dimensions then previously 
The closest prior work is ref.~\cite{Kantian2019}, which tackled an inherently different problem only achieving  a factor of 2 lower in bond dimension. 
We also compare multiple different algorithms for performing tensor contractions as well as contrast two qualitatively different types of physical systems.  Furthermore, the approach between the two works is different. All prior memory-distributed work has used a matrix/vector formalism for distributed computing while we work directly distributing higher-order tensors including throughout the intermediate steps of optimization. 

\section{Algorithms}\label{sec:alg}
\begin{table*}[]
    \centering
    %\resizebox{\textwidth}{!}{ %make table fit without guessing
    % \begin{tabular}{ll|ll|l|l}
%  & \multicolumn{1}{l|}{} & \multicolumn{2}{c|}{Memory } & \multicolumn{2}{c}{BSP cost for Davidson iteration}\tabularnewline
% Algorithm & Flops  & Davidson ($M_D$) & Environments & BSP supersteps & BSP comm cost \tabularnewline
% \hline 
% \hline 
% List & \multirow{2}{*}{$O((m/q)^{3}kd^{2})$} & \multirow{2}{*}{$O((m/q)^{2}kd^{2})$} & \multirow{3}{*}{$O(N(m/q)^{2}k)$} & $O(N_b)$ & $O(M_D/p^{2/3})$ \tabularnewline
% Sparse-Sparse &  &  & & $O(1)$ & $O(M_D/p^{1/2})$ \tabularnewline
% Sparse-Dense & $O(m^{3}kd^{2})$ & $O(m^{2}kd^{2})$&  & $O(1)$ &$O(M_D/p^{1/2})$ \tabularnewline
% \end{tabular}

\begin{tabular}{ll|ll|ll}
 & \multicolumn{1}{l|}{} & \multicolumn{2}{c|}{Memory} & \multicolumn{2}{l}{BSP cost for Davidson iteration}\tabularnewline
Algorithm & Flops & Davidson($M_{D})$ & Environments & BSP supersteps & BSP comm cost\tabularnewline
\hline 
\hline 
List & $O((m/q)^{3}kd^{2})$ & $O((m/q)^{2}kd^{2})$ & \multirow{3}{*}{$O(N(m/q)^{2}k)$} & $O(N_{b})$ & $O(M_{D}/p^{2/3})$\tabularnewline
Sparse-Sparse & $O((m/q)^{3}kd^{2})$ & $O((m/q)^{2}kd^{2})$ &  & $O(1)$ & $O(M_{D}/p^{1/2})$\tabularnewline
Sparse-Dense & $O(m^{3}kd^{2})$ & $O(m^{2}kd^{2})$ &  & $O(1)$ & $O(M_{D}/p^{1/2})$\tabularnewline
\end{tabular}
    %}
    \caption{Complexity of each algorithm implementation, in terms of number of sites $N$, bond dimension $m=\sum^{N_b}_\ell b_\ell$ and number of blocks $N_b$, MPO bond dimension $k$, and physical dimension $d$. Here we are using a empirically motivated model where the $\ell$th block has auxiliary dimension $b_\ell=\lfloor (m/q) r^\ell\rfloor$; the values $q=4, r=0.6$ for spins and $q=10, r=0.65$ for electrons are rough estimates of the parameters fit to our data. }
    \label{tab:alg_complexity}
\end{table*}

\begin{algorithm}[t]

\centering

    \begin{algorithmic}[1]
    \small
    \State{\textbf{Input: } Tensor objects A and B containing lists of quantum number blocks to represent tensors  $\tsr A\in\mathbb{R}^{s_1\times\cdots\times s_{r_A}}$ and  $\tsr B\in\mathbb{R}^{d_1\times\cdots\times d_{r_B}}$, contracted index/indices \texttt{mid} }
    \State{$\vcr i_A$ = getIndexLocations(A, \texttt{mid}) }
    \State{$\vcr i_B$ = getIndexLocations(B, \texttt{mid}) }
    \State{C = new Tensor(order = $r_A+r_B-\vcr i_A$.size())}
    \State{CBlocks = [], qToBlockIndex =\{\} }
    \For{ABlock in A.blocks}
        \State{$\vcr q_A$ =  getQuantumNumberLabels(ABlock)}
        \For{BBlock in B.blocks}
             \State{$\vcr q_B $=  getQuantumNumberLabels(ABlock)}
              \If{$\vcr q_A[i_A] \not= \vcr q_B[i_B]$}
              continue
             \EndIf
          \State{$\vcr q_C$ = []} \Comment{build $\vcr q_C$ from remaining labels}
          \For{i$\gets$0,\dots,$\vcr q_A$.size()}
            \If{i not in $\vcr i_A$} $\vcr q_C$.append($\vcr q_A[i]$)
            \EndIf
          \EndFor 
          \For{i$\gets$0,\dots,$\vcr q_B$.size()}
            \If{i not in $\vcr i_B$} $\vcr q_C$.append($\vcr q_B[i]$)
            \EndIf
          \EndFor 
            \If{$\vcr q_C$ is in  qToBlockIndex}
            \State{Cidx = qToBlockIndex[$\vcr q_C$]}
            \State{CBlocks[Cidx] += ABlock.contract(BBlock) } \Comment{CTF}
            \Else
            \State{CBlocks.append(ABlock.contract(BBlock)) } \Comment{CTF}
            \State{qToBlockIndex[$\vcr q_C$] = CBlocks.size()} 
            \EndIf
              %\EndIf
        \EndFor
    \EndFor
\end{algorithmic}
    \caption{Contraction of two tensor objects $A$ and $B$ to resultant tensor object $C$, each composed of a list of quantum number blocks}
    \label{alg:QNBlockContract}
\end{algorithm}
%\end{figure}

Our parallelization\footnote{source code can be found at \url{https://github.com/ClarkResearchGroup/tensor-tools}} is developed on top of the serial DMRG code \TT~and 
builds upon the Cyclops Tensor Framework \cite{solomonik2014ctf} to leverage large scale computing resources. Cyclops handles the shared memory structure for the tensor, contraction, and provides a pass through to parallel linear algebra routines, \textit{e.g.} SVD, from ScaLAPACK and the High-Performance Tensor Transpose library (HPTT)\cite{hptt2017}. A variety of alternative tensor libraries exist, including ones focused on sparse tensor contractions~\cite{Kats_sp_tensor2013,chou2018formats,kjolstad2019workspaces,libtensor} or distributed dense contractions~\cite{doi:10.1021/jp034596z,lai2013framework,springerlink_ga1,mutlu2019toward} as Cyclops, as well as parallel block-sparse tensor contractions~\cite{calvin2015scalable,peng2016massively,libtensor}.
Development of DMRG using the last category (block-sparse libraries) would be promising future work, as they may more effectively exploit the sparsity structure of tensors we consider.
However, their use may entail potential overheads both in implementation complexity, e.g., unlike Cyclops, DBCSR requires manual specification of processor grids~\cite{sivkov2019dbcsr}, as well as performance.
Block-sparse tensor contraction libraries~\cite{calvin2015scalable,peng2016massively,libtensor} have been previously only used for electronic structure methods with workloads that differ significantly from DMRG.

Our focus in this work is in the limit of large bond dimension where serial codes take significant time and the tensors become large enough that the gains of parallelization outweigh its overhead. DMRG, at fixed bond dimension, has linear scaling with system size and our parallelization approach preserves this scaling.
\subsection{Tensor Structure}
\begin{figure}
\centering
\includegraphics[width=\columnwidth,trim=2cm 0cm 2cm 0cm, clip]{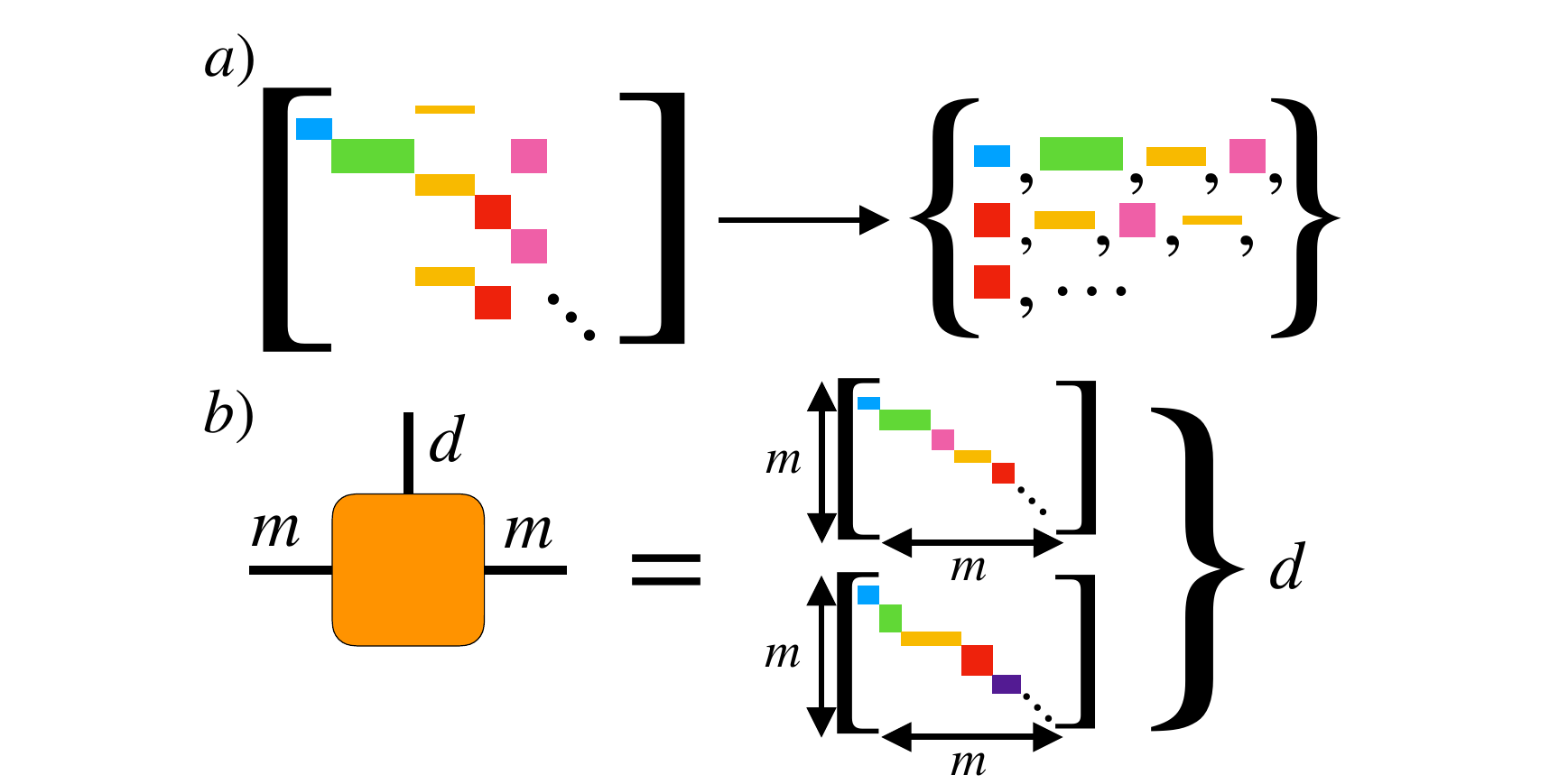}
\caption{Cartoon of the tensor quantum block structure. \textit{a)} For the list format, we deconstruct the many blocks into a list of distributed memory tensors. \textit{b)} each tensor in an MPS has a special block diagonal structure, where the blocks form $d$ $m\times m$ matrices.  }
\label{fig:structure_cartoon}
\end{figure}
\begin{figure}[ht]
  \centering
    \subfloat[$J_1-J_2$ 20x10 Cylinder]{
  \centering
 \includegraphics[width=0.5\columnwidth]{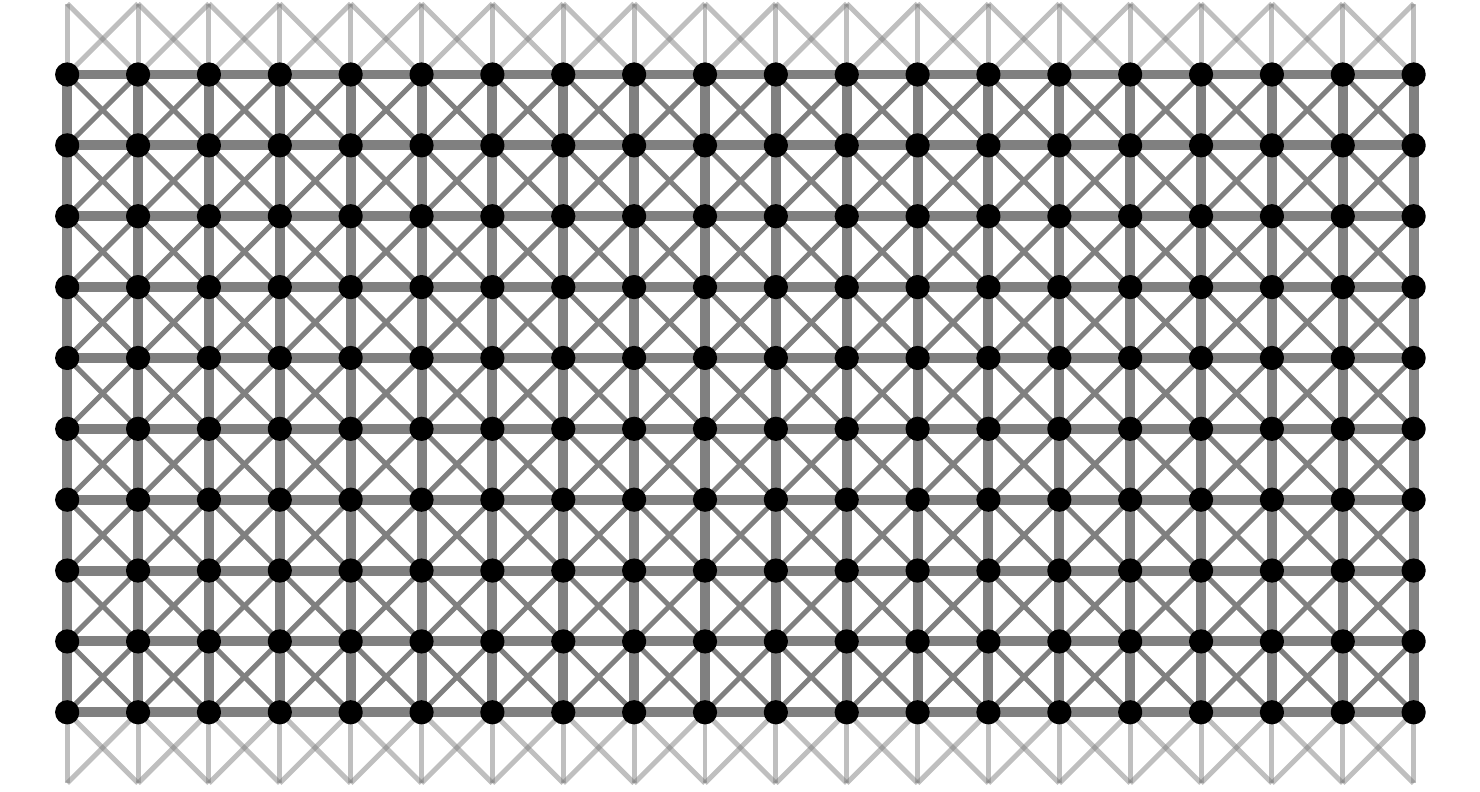}
  }
  \subfloat[\label{fig:TriLattice} 6x6 triangular Cylinder (XC6)]{
  \resizebox{!}{65pt}{\includegraphics[width=0.5\columnwidth]{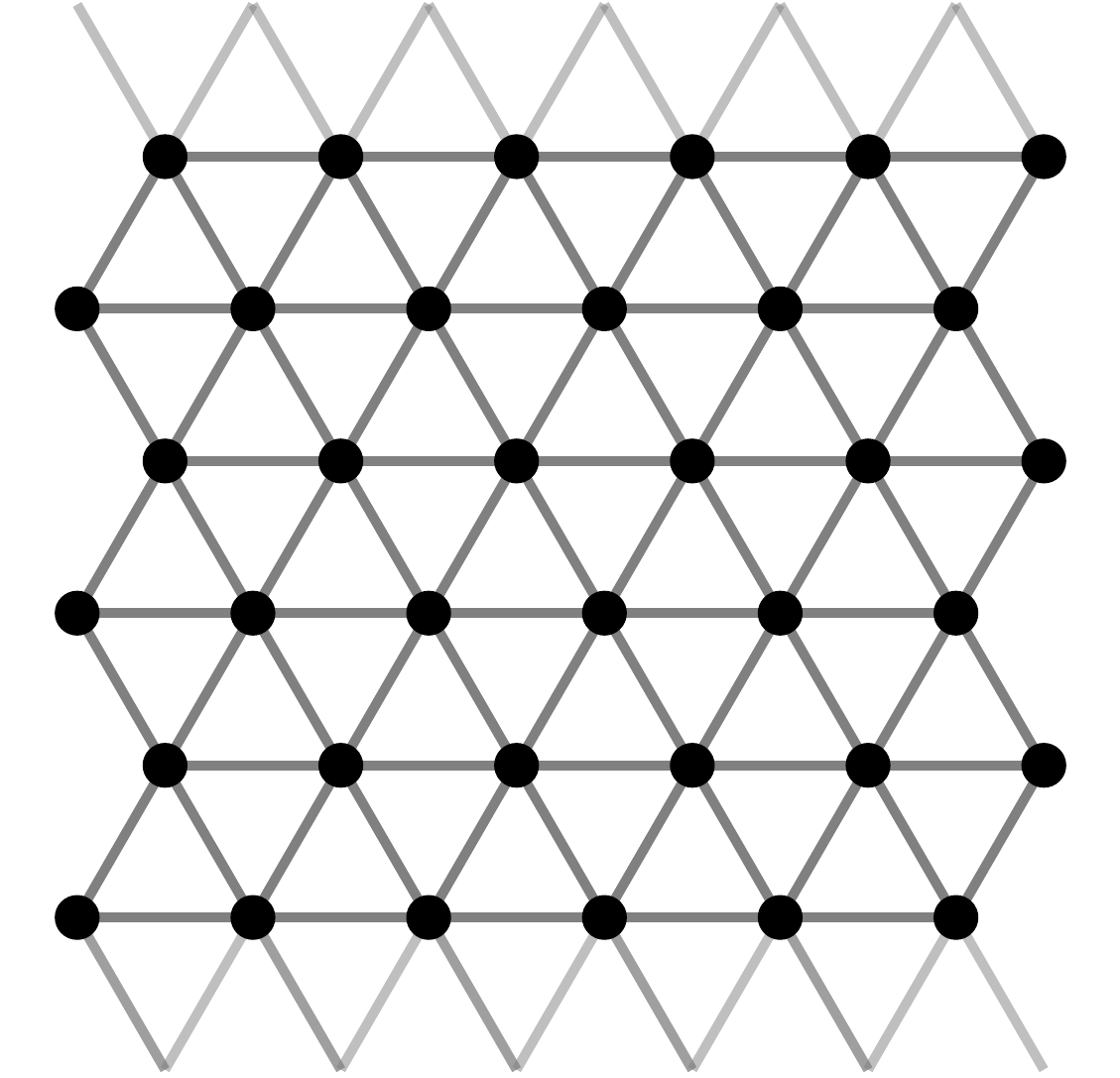}}
  }
  \caption{Lattice structure of the two benchmark systems. }\label{fig:lattices}\end{figure}
Our tensors have block-sparsity due to the quantum number structure of the matrix product state.  We consider and test three different interfaces for representing this block-sparsity in memory which correspond to three different algorithms for contraction. 
The performance of each interface is dependent upon the structure of the quantum number blocks. 
\\

\noindent \textit{list} algorithm: The first algorithm for tensor contraction, the \textit{list} algorithm, stores each tensor as a set of memory distributed tensor blocks $\tsr T_{q^{(\ell)}}$ for each tuple of quantum number labels $q^{(\ell)}$, initialized via a dense Cyclops tensor (cartoon shown in fig.~\ref{fig:structure_cartoon}a). 
To contract two tensors, the quantum number label structure must be analyzed to determine which blocks contract as well as the resulting block structure, shown in algorithm~\ref{alg:QNBlockContract}. 
All possible combination of blocks that have the same label along contracted indices are contracted together via parallel tensor contractions leaving the remaining labels that are not among the contracted indices as the resultant labels.
The resultant tensor is then made up of a new set of distributed tensor blocks. \\

\noindent \textit{sparse-dense}: Alternatively, tensors can be formed by combining all blocks into a single distributed tensor.  To form the tensor, each quantum number label is mapped to a unique tensor index range of dimension of the quantum number label.  Note that this tensor has non-trivial sparsity.  
All operations such as addition and contraction will produce the same output as with the list format, with a single contraction call. 
Tensors stored in this format have a higher memory cost, so that each MPS tensor now has storage cost $dm^2$, the same as without quantum numbers. 
To conserve memory but exploit dense-dense tensor contraction performance, environment tensors, MPS, and MPO tensors are stored in a sparse format. Intermediate tensors of the Davidson routine are stored as dense, which we call the \textit{sparse-dense} algorithm.  \\

\noindent \textit{sparse-sparse}: When a system has a quantum number with many relevant labels (e.g. particle number) or there are many combination of labels (e.g. two or more conserved quantities), we observe that the single tensor of blocks is quite sparse as shown in fig.~\ref{fig:sparsity}. 
Therefore we can store all intermediate tensors in a Cyclops sparse tensor, or the \textit{sparse-sparse} algorithm. 
This has an additional overhead of keeping track of non-zero elements of each tensor along with determining output sparsity and the distributed distribution of elements. 
Knowledge of quantum number labels allows for pre-computation of the output sparsity, which can be provided to Cyclops to control memory consumption during contraction. \\

For all algorithms, the SVD portion of DMRG is performed via the list method. 
For sparse-sparse and sparse-dense this requires that the blocks are extracted from the single tensor and put into a temporary list format. 
As SVD is only defined on a order-2 tensor, the tensor indices are `wrapped' to form an effective order-2 matrix with a row index and a column index. 
Quantum numbers of the singular vector `tensors' are used to calculate legal quantum numbers for the virtual index between the vectors and the singular values. Once this has been computed, a subset of blocks are reshaped into a matrix, grouped via similar quantum numbers along a row or column index, and decomposed. This produces a block structured tensor which can be reshaped into two order-3 tensors, and singular values absorbed into either the left or right tensor, along the direction of the sweep (see fig.~\ref{fig:dmrg_diag}e). 
We utilize a distributed SVD routine through ScaLAPACK \cite{blackford1997scalapack}, so as to minimize redistribution costs of moving data onto a single node to call a serial SVD routine. 
Finally any sparse structure is recovered by rearranging data in parallel from the list format into a single sparse tensor.

We include in table \ref{tab:alg_complexity} complexity of each format using a simplified, empirically motivated block structure model.
We also provide communication cost models for the most costly contractions in the method, which are quantified using the Bulk Synchronous Parallel (BSP)~\cite{valiant1990bridging,skillicorn1997questions}, in terms of supersteps (number of global synchronizations) and communication cost (amount of data sent along the critical path of execution).
The costs are based on the algorithms used by Cyclops, which have a cost that depends on available memory~\cite{solomonik2014ctf,Solomonik:2017:SBC:3126908.3126971}.
We assume that enough memory is available to achieve the minimal possible communication when executing a block-wise contraction with all processors, but that no (at most a constant factor of) additional memory is available when all blocks are multiplied within a single tensor contractions.
The analysis demonstrates that the choice of best method depends on the problem parameters (e.g., number of blocks), as there is a trade-off between synchronization and communication costs.

\section{Physical Systems} \label{sec:systems}
Condensed matter systems fall broadly into two classes:  electron systems for which the underlying itinerancy of the electron matters and spin-systems where the spins can be treated as stationary but interact with each other.  We benchmark two challenging systems (one in each category) which have been heavily studied but for which there is not yet consensus about the underlying physics. The first, a spin system (called \textit{spin} throughout the text) with $d=2$ physical degrees of freedom, is the $J_1-J_2$ Heisenberg model,
\begin{align*}
    H = J_1 \sum_{\langle i,j \rangle} \bm{S}_i \cdot \bm{S}_j + J_2\sum_{\langle \langle i,j \rangle\rangle} \bm{S}_i \cdot \bm{S}_j,
\end{align*}
where $\bm{S}_i$ is a spin operator on site $i$, while $\langle i, j \rangle$ and $\langle \langle i,j \rangle \rangle$ iterate over sites on a 2D lattice that are, respectively, directly and diagonally adjacent. At $J_2/J_1=0.5$ there is disagreement about the phase coming both from multiple DMRG studies \cite{Jiang2012,Gong2014,Wang2018} as well as other tensor network approaches \cite{Haghshenas2018,Liu2018,PhysRevB.96.014414}. 
While we will not resolve that problem here, using microbenchmarks we will show that we can reach a bond dimension $m$ significantly above current state-of-the-art simulations at comparable efficiency and significantly reduced wall-clock time.

\begin{figure}
    \centering
    \includegraphics[width=\columnwidth]{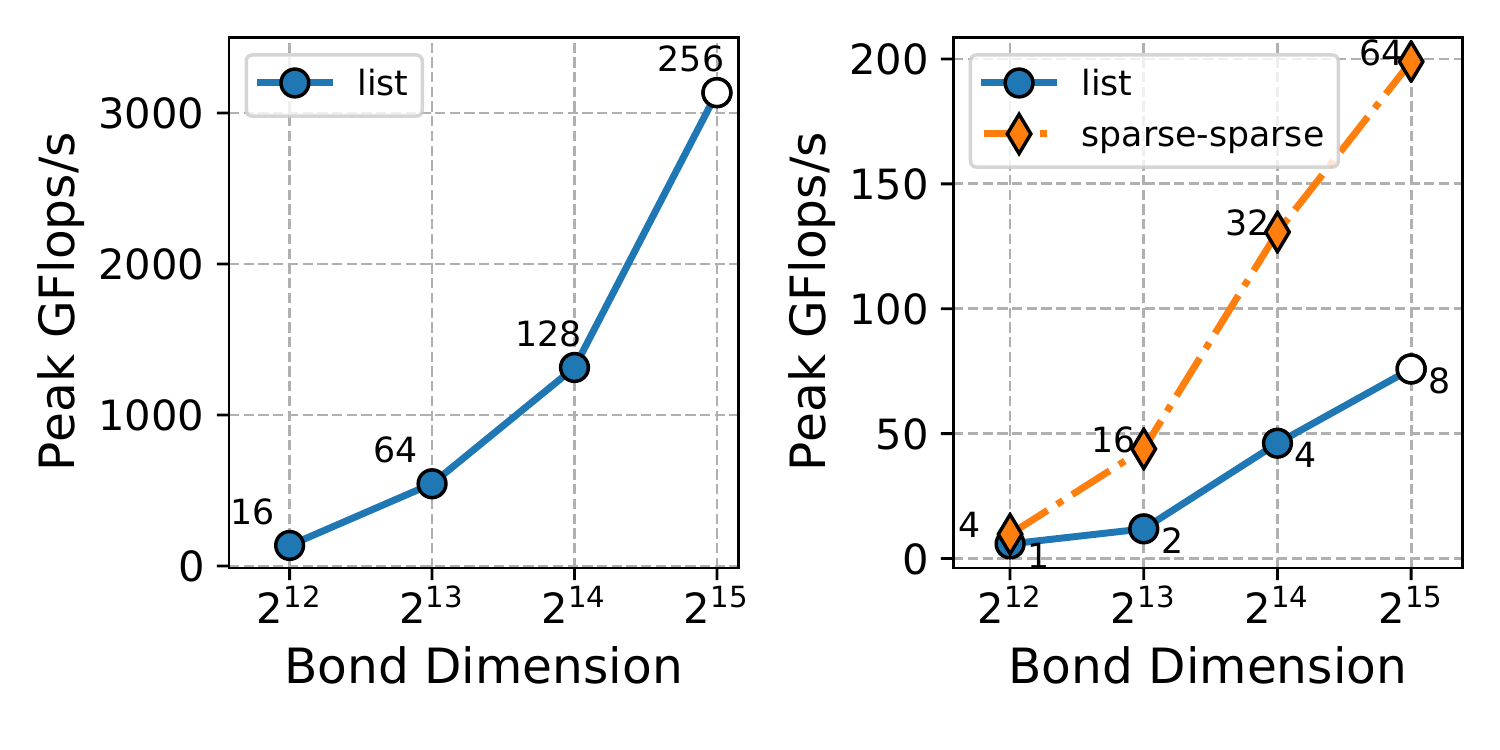}
    \caption{Summary of the peak performance of the spin (\textit{left}) and electron (\textit{right}) systems. GFlops/s are averaged over a sweep of 10 (1) sites for the spin (electron) system respectively. The largest spin and electron list result, denoted with an open marker, is estimated from half sweep data. Annotated are the number of nodes used.}
    \label{fig:BWFlops}
\end{figure}
In addition to looking at the spin system above, we also benchmark an electron system (called \textit{electron}) with $d=4$ physical degrees of freedom, the triangular Hubbard model,
\begin{align*}
    H = -t \sum_{\langle i,j \rangle,\sigma} \left( c^\dagger_{i\sigma} c_{j\sigma} + h.c.\right) + U\sum_i n_{i\uparrow}n_{i\downarrow},
\end{align*}
where $c^{\dagger}_{i\sigma}$ is an electron creation operator at site $i$ with spin $\sigma=\left\{\uparrow,\downarrow\right\}$ and $n_{i\sigma} = c^\dagger_{i\sigma}c_{i\sigma}$ is the number operator of electrons with spin $\sigma$ on site $i$. We set $t=1$ and $U=8.5$ and use $N$ electrons with $N_\uparrow=N_\downarrow=N/2$.  The Hubbard model simulations differ qualitatively from the Heisenberg simulations both in their physical content (being itinerant electrons instead of stationary spins) as well as their tensor structure.  In particular, the physical degrees of freedom now contain four states per site (i.e $d=4$) and, more importantly, there are two conserved global symmetries, spin and particle number. These two symmetries translate into two quantum numbers per label which significantly increases both the number of blocks and sparsity of blocks for the same bond dimension (see fig.~\ref{fig:block_sizes} and fig.~\ref{fig:sparsity}).  Different techniques also disagree on the phase diagram of this model \cite{Shirakawa2017,Szasz2018}.
\begin{figure}
    \centering
    \includegraphics[width=\columnwidth]{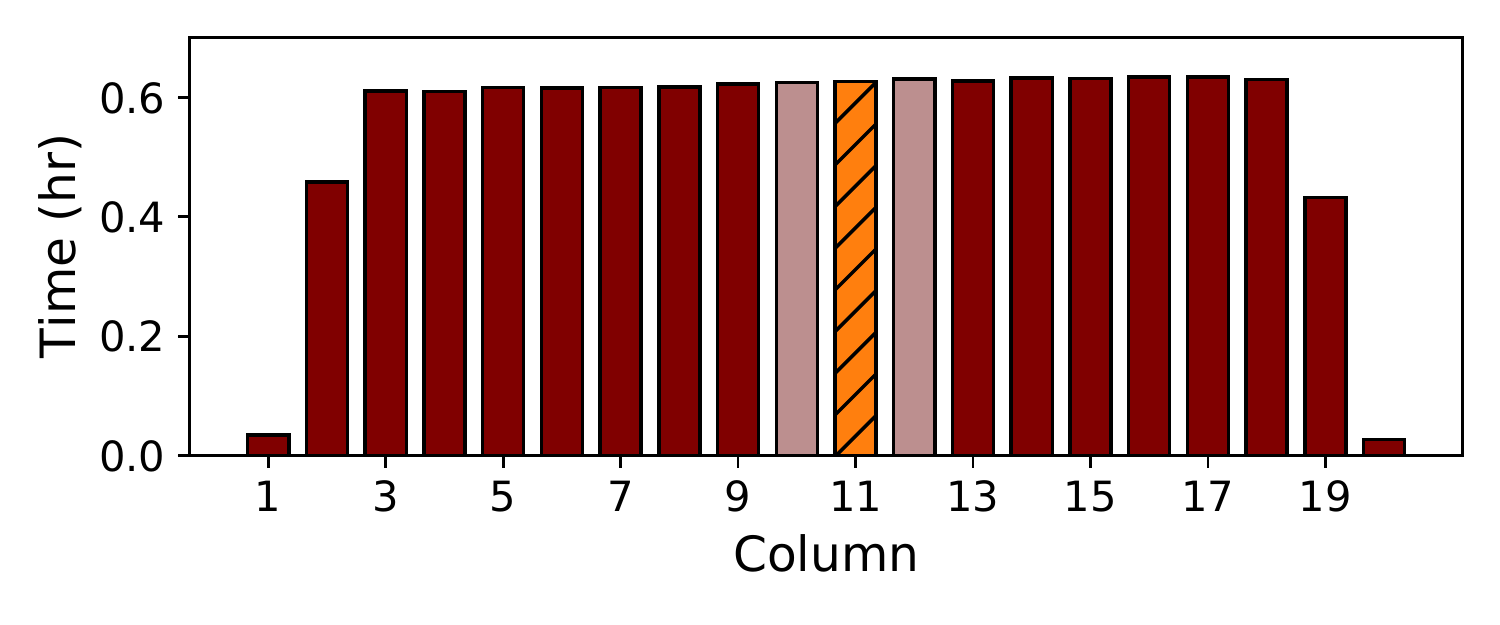}
    \caption{Time spent for each column of 10 sites for a full sweep at bond dimension $m=8192$ for list spins. Spin benchmark optimizations are carried out on the middle 3 columns, denoted in light maroon and orange, which share similar timings to all non-edge columns in a full sweep. We report timing for only the hatched center column to ignore any additional edge effects. }
    \label{fig:columnComparison}
\end{figure}

The Hamiltonian for both our systems is encoded as a MPO.   The structure of this MPO is not unique.  Because we are using ITensor to compare against, to ensure equity in this comparison, we use exactly the same MPO ITensor generates by directly using their AutoMPO functionality \cite{itensor}. 
\section{Numerical Experiments}\label{sec:exp}
Numerical experiments are performed on Blue Waters and Stampede2 supercomputers. All reported Blue Waters timings use the Cray XE6 nodes with 64 GB of RAM and dual 8-``core'' processors per node connected via Cray's Gemini interconnect, while with Stampede2 we utilize only Knight's Landing (KNL) nodes which have a 68 core processor, 96 GB of DDR4 RAM, 16 GB of MCDRAM, connected via an Intel Omni-Path interconnect. For all Stampede2 data and sparse algorithm data we use Intel's MKL library which includes BLAS, batched BLAS, ScaLAPACK,  and Sparse-Sparse routines. On Blue Waters for the list algorithm, we use Cray's LibSci library for BLAS and ScaLAPACK routines.    

In order to mitigate differences in MPS block sizes and distributions, we developed an interface to convert ITensor MPS data to a readable format for Cyclops. This enables us to have extremely close comparison to single node performance for all available problem sizes with comparable single node performance. 

To calculate the number of flops, we measure FLOP operations using the built in Cyclops routines for the list method. 
The resulting measurement is then used as basis for ITensor, list, and sparse method performance rate calculations (flops/s). The peak GFlops/s performance rate is shown in fig.~\ref{fig:BWFlops}, where we obtain a maximum performance of $3.1$ TFlops/s on Blue Waters and 198 GFlops/s on Stampede2.
\begin{figure}
    \centering
    \includegraphics[width=\columnwidth]{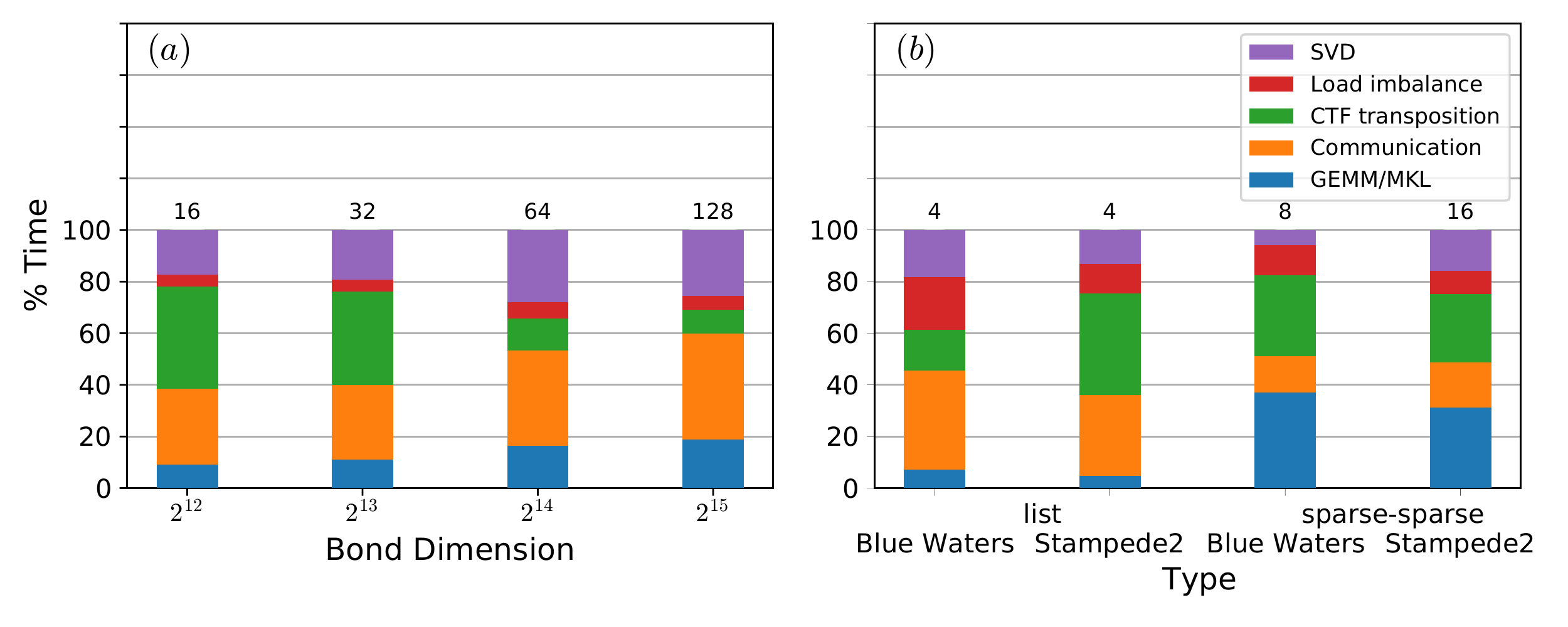}
    \caption{Breakdown of percentage of time spent for (a) spins on Blue Waters with 16 MPI processes/node and (b) electron systems at $m=2^{14}$ on both Blue Waters (16 processes/node) and Stampede2 (64 processes/node). Annotated are the number of nodes used. Communication costs include MPI calls excluding those in SVD (ScaLAPACK \texttt{pdgesvd}); CTF transposition includes CTF mapping, transpose operations, and other small serial operations; and load imbalance is measured via \texttt{MPI\_Barrier()} calls.}
    \label{fig:commCosts}
\end{figure}

In order to compare the single node performance of a shared memory architecture application, we benchmark ITensor on a single node with 32 threads/node on Blue Waters and Stampede2. ITensor takes advantage of threaded BLAS and LAPACK routines; on Blue Waters we use the Cray LibSci library and on Stampede2 the Intel MKL library. To approximate timings for larger problem sizes than can fit on one node, we take the maximum performance rate (GFlops/s) and report the extrapolated numbers. On Blue Waters we use the Cray XE6 high memory nodes with 128 GB of RAM.

A half-sweep of a DMRG algorithm performs $N$ optimizations (the other half-sweep optimizes sites in the opposite order). A typical MPS (in canonical form) has little variance in bond dimension across the system (modulo those near the edge).  Therefore, the time per site should be largely uniform outside the first few and last sites; we validate this in fig.~\ref{fig:columnComparison} comparing the time of each column (10 sites) over an entire sweep.   For the spins system, instead of timing all sites, we therefore optimize the middle 3 columns (or 30 sites) of the lattice reporting the timing of the middle column; toward that end, we ensure the MPS has the reported bond dimension on those 30 sites but not necessarily outside of it.  This is often done by an untimed sweep where we grow the bond dimension.   All sweep times and GFlops/s reported for the Heisenberg model are from the middle 10 sites of these 30 site sweeps unless noted. For $m<8192$ we additionally have an SVD cutoff of $10^{-9}$ and $m\geq8192$ a cutoff of 
$10^{-12}$.  For the Hubbard model (electrons), we choose to benchmark a single DMRG step (the 15th and 16th sites) rather than the entire system. 

\subsection{Spins ($d=2$)}
\begin{figure*}[h]

    \centering
    \subfloat[DMRG Relative Efficiency\label{fig:bw_spins_weak}]{\includegraphics[width=0.9\columnwidth]{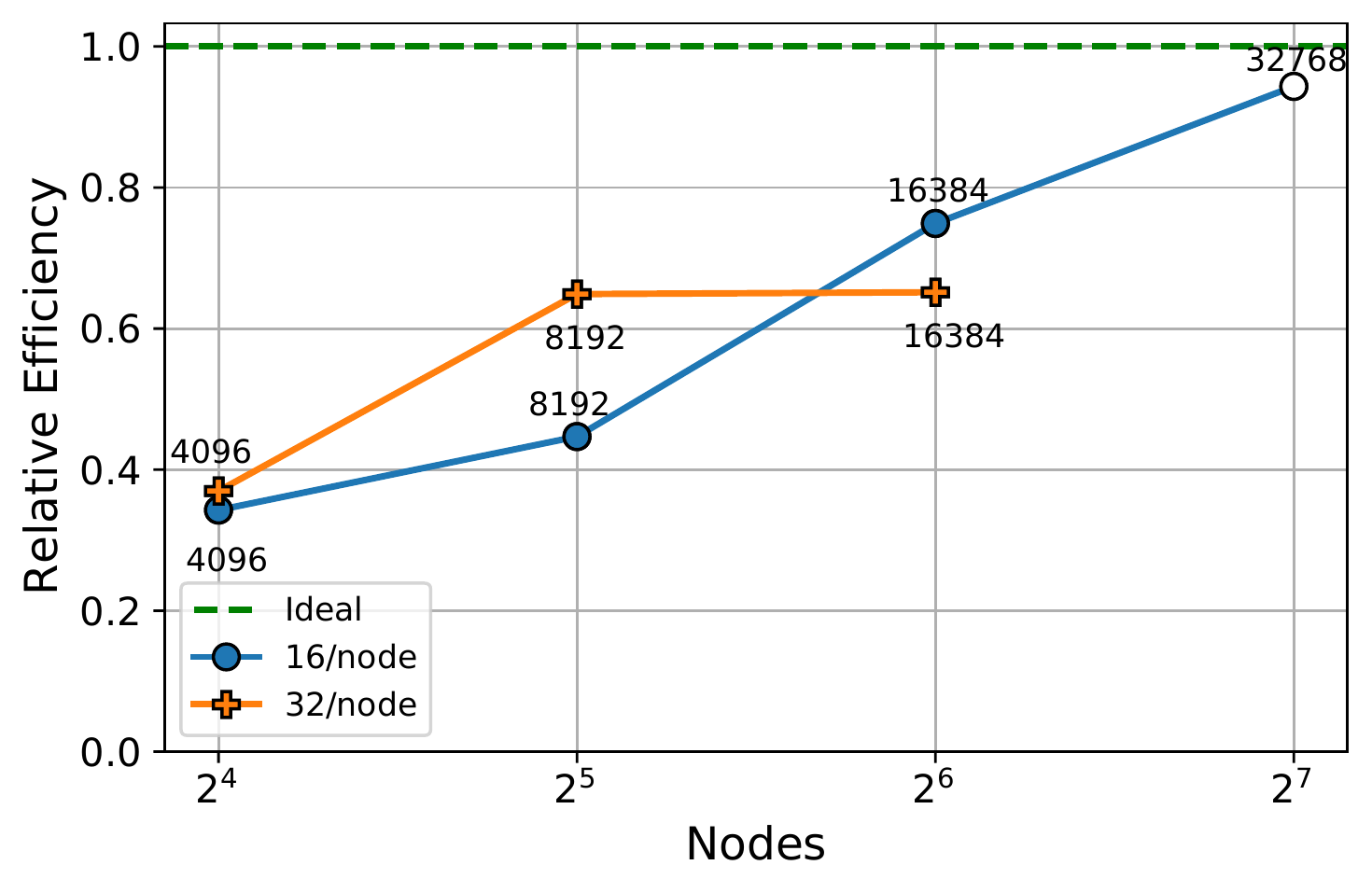}}
    \subfloat[Peak Relative Efficiency\label{fig:spins_peak_eff}]{\includegraphics[width=0.9\columnwidth]{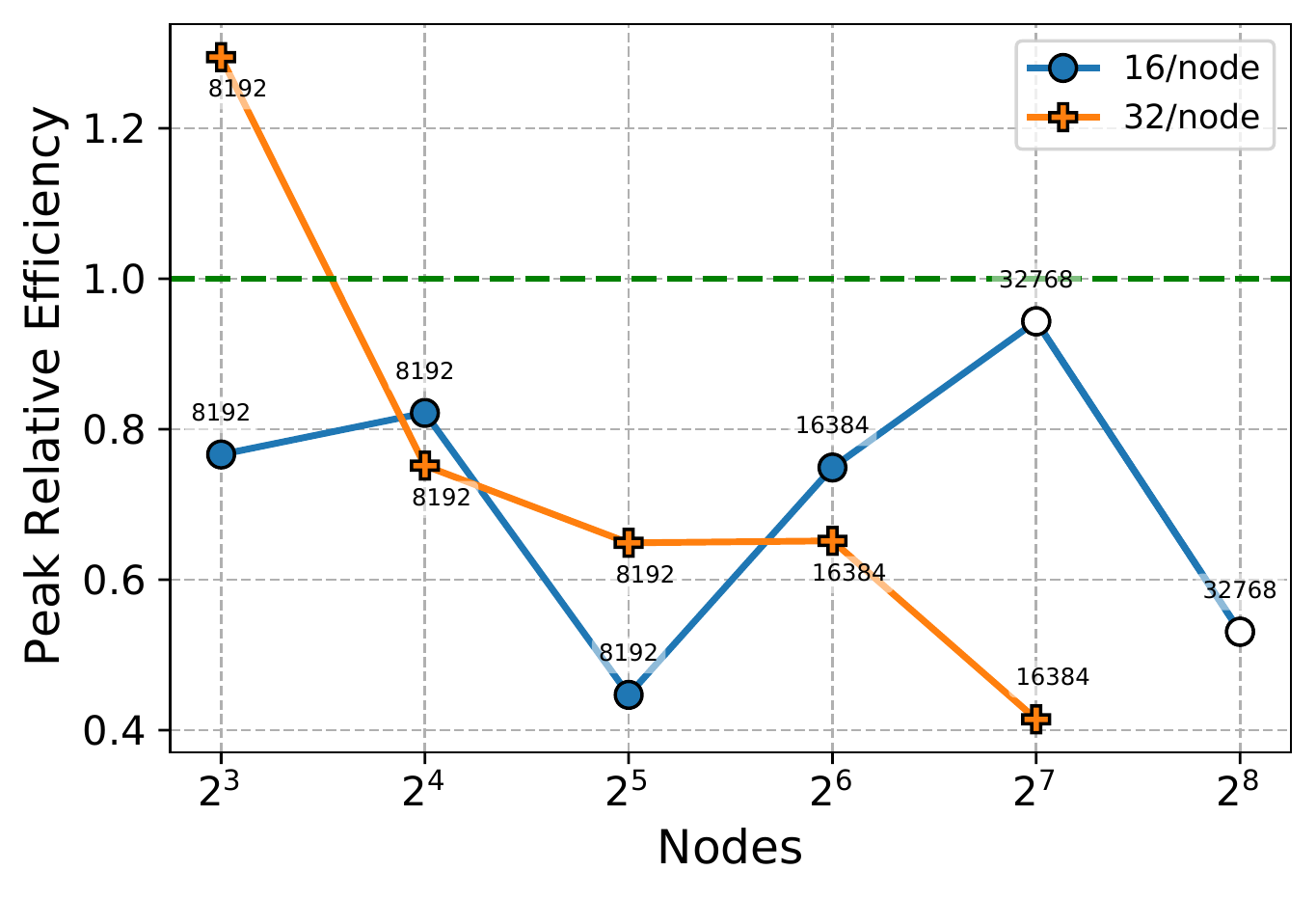}}
    \caption{Weak scaling using  fixed $m/\text{node}$ (\textit{Left}) and peak relative efficiency with respect to single node ITensor (\textit{Right}) for list spins on Blue Waters. Bond Dimensions $m$ are annotated on each point; relative efficiency is calculated by relative GFlops/s/node compared to ITensor on a single node at $m=4096$. Peak relative efficiency is taken as the highest relative efficiency measured at a given node count.  Open fill points are extrapolated from one half sweep.}
    \label{fig:WeakScalingSpins}
\end{figure*}

\begin{figure}
    \centering
    \includegraphics[width=\columnwidth]{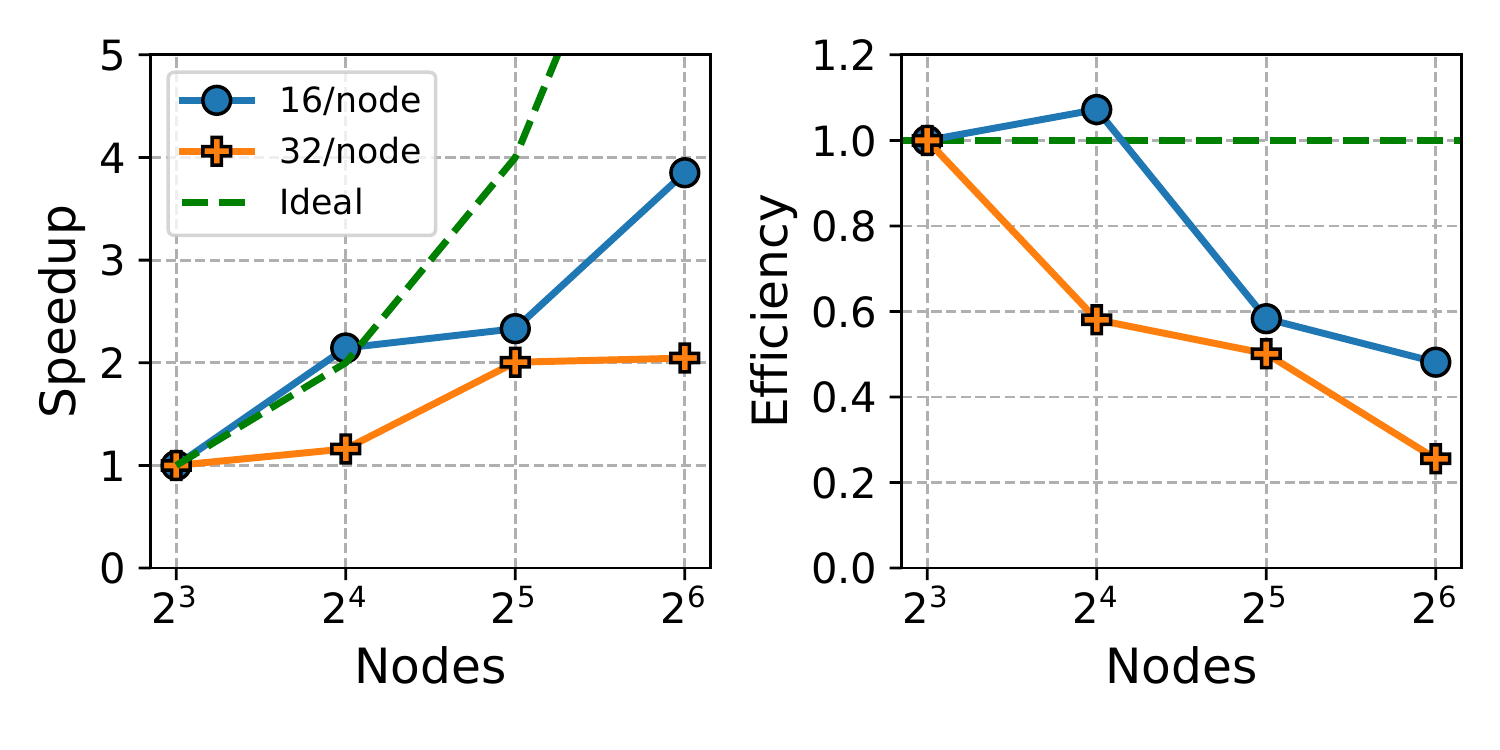}
    \caption{Strong scaling speedup (\textit{left}) and efficiency (\textit{right}) of the list format spin system at $m=8192$ on Blue Waters. }
    \label{fig:bw_spins_strong}
\end{figure}

\begin{figure}
    \centering
    \includegraphics[width=\columnwidth]{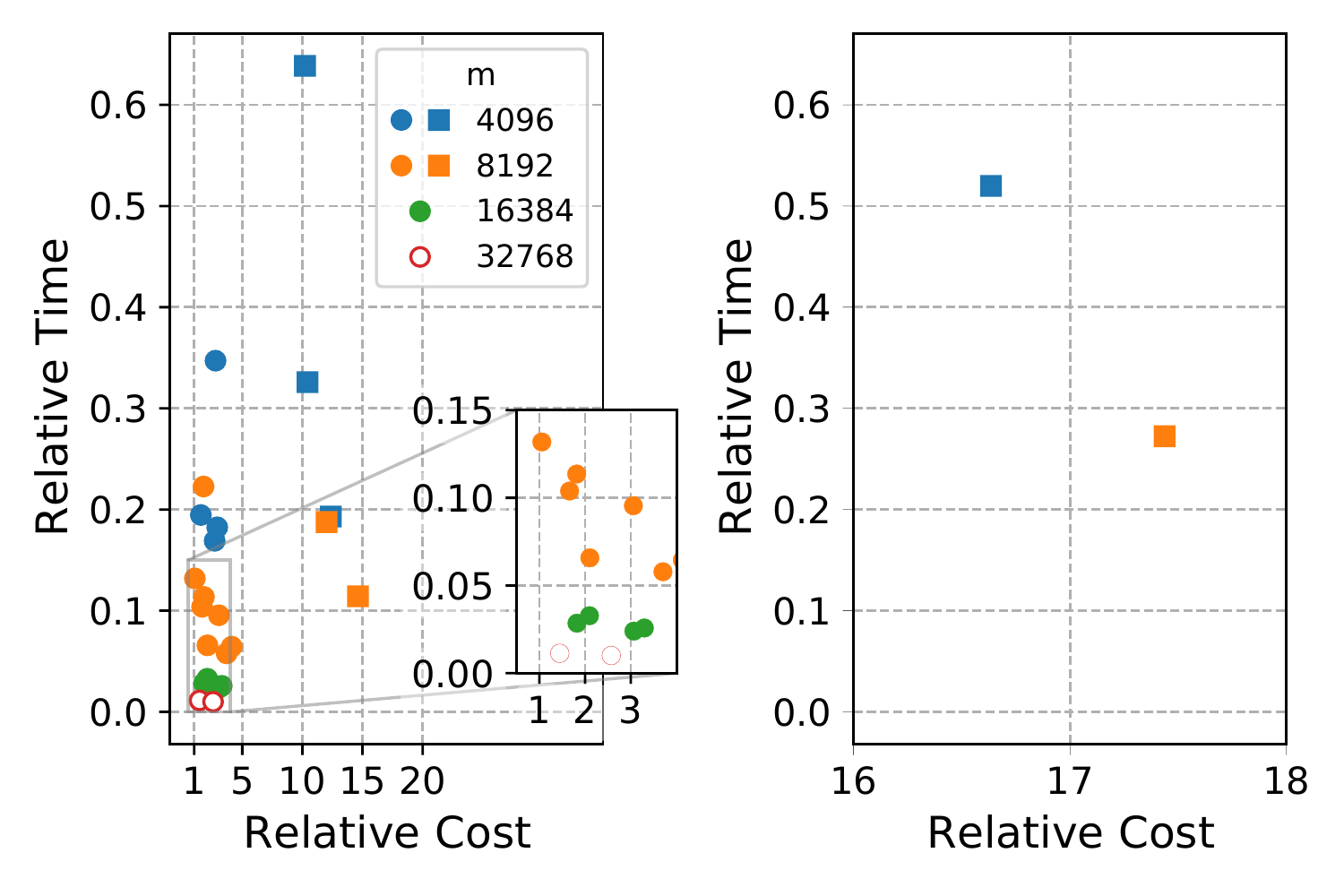}
    \caption{Spin system execution time and node hour cost relative to single node ITensor maximum performance rate (at $m=4096$) using lists (circles) and sparse-dense (squares) on Blue Waters \textit{left} and Stampede2 \textit{right} by varying hyperparameters (node count and MPI processes/node); open points are extrapolated from a half sweep. }
    \label{fig:spins_time_cost}
\end{figure}

 On Blue Waters, where single node throughput is (comparatively) low, 
 in figs~\ref{fig:WeakScalingSpins},~\ref{fig:bw_spins_strong} and~\ref{fig:spins_time_cost},
 we observe that list method is the fastest and most cost effective approach, performing significantly better than sparse-dense.  We consider both the strong and weak scaling of the list method on Blue Waters.  In looking at this scaling, the relevant `problem size' to increase is the bond dimension, which dictates the accuracy of the DMRG approximation.  We find in the strong scaling that the  efficiency and speedup stays ideal only for a modest increase in the number of nodes (i.e. going from $2^3$ to $2^4$ nodes).  Beyond this, we do not see significant gains in increasing the number of nodes at fixed bond dimension (see fig.~\ref{fig:bw_spins_strong}) with an efficiency falling to approximately 60\% under an additional doubling of the nodes. 

Heuristically, after determining the maximum problem size on a single node, we observe in fig.~\ref{fig:bw_spins_weak}
that doubling the number of nodes when doubling the bond dimension maintains good efficiency.
 We measure efficiency relative to single node execution of ITensor.
 We obtain near ideal efficiency at the largest node count, with bond dimension $m=32768$. Note that doubling problem size does not double work but increases work/node by a factor of 8 and increases memory/node by a factor of 4. 
 
 As the problem size grows, the list algorithm is able to more efficiently use the computational resources compared to that of ITensor, despite growing communication costs. In our performance breakdown in fig.~\ref{fig:commCosts}(a) this is supported as when bond dimension is increased, there is a corresponding increase in local matrix-matrix multiplication (GEMM), signifying improvement in efficiency.
 A significant but not a dominant amount of time is spent in SVD and communication, which are expected parallel bottlenecks.
We also consider peak relative efficiency in fig.~\ref{fig:spins_peak_eff}
finding good peak performance both at small and large node counts, with a dependence on MPI processes per node which crossovers at $n=2^6$ between $32/\text{node}$ and $16/\text{node}$ being preferred. 

Finally, in fig.~\ref{fig:spins_time_cost} we consider the relative time and cost of a variety of node counts, processes per node, bond dimensions, and algorithms.  We find that we can achieve speedups of 5.9X to 99X of the wall clock time, as bond dimensions grow from $m=4096$ to $m=32768$, at a relative node cost that is 1.5 times that of running ITensor.  In order to theoretically compare to problems which do not fit on a single node, we calculate times using the maximum single node performance rate. We find that by considering the Pareto optimal curve in fig.~\ref{fig:spins_time_cost}, which selects the best/minimum relative time for a given cost and bond dimension, on Blue Waters the curve entirely consists of list algorithm simulations. 
 
\subsection{Electrons ($d=4$)}
In this section, we now turn to the 6x6 triangular Hubbard Model with the larger local dimension, $d=4$.  The additional quantum number symmetry --- particle number ---  
implies significantly more blocks and hence greater sparsity (see fig.~\ref{fig:block_sizes}).  Due to the lower sparsity we benchmark the sparse-sparse tensor algorithm in addition to the list method.  Although these both should cost roughly the same number of total flops, the overhead is different in list, we serially enumerate over quantum blocks contracting them; this has an overhead coming from contracting small tensors in a distributed way.  On the other hand, the sparse-sparse approach contracts a single pair of tensors;  unfortunately, extra overhead exists from sparse operations.

\begin{figure*}[ht]
    \centering
    \subfloat[DMRG Weak Scaling\label{fig:bw_fermion_weak}]{\includegraphics[width=1.2\columnwidth]{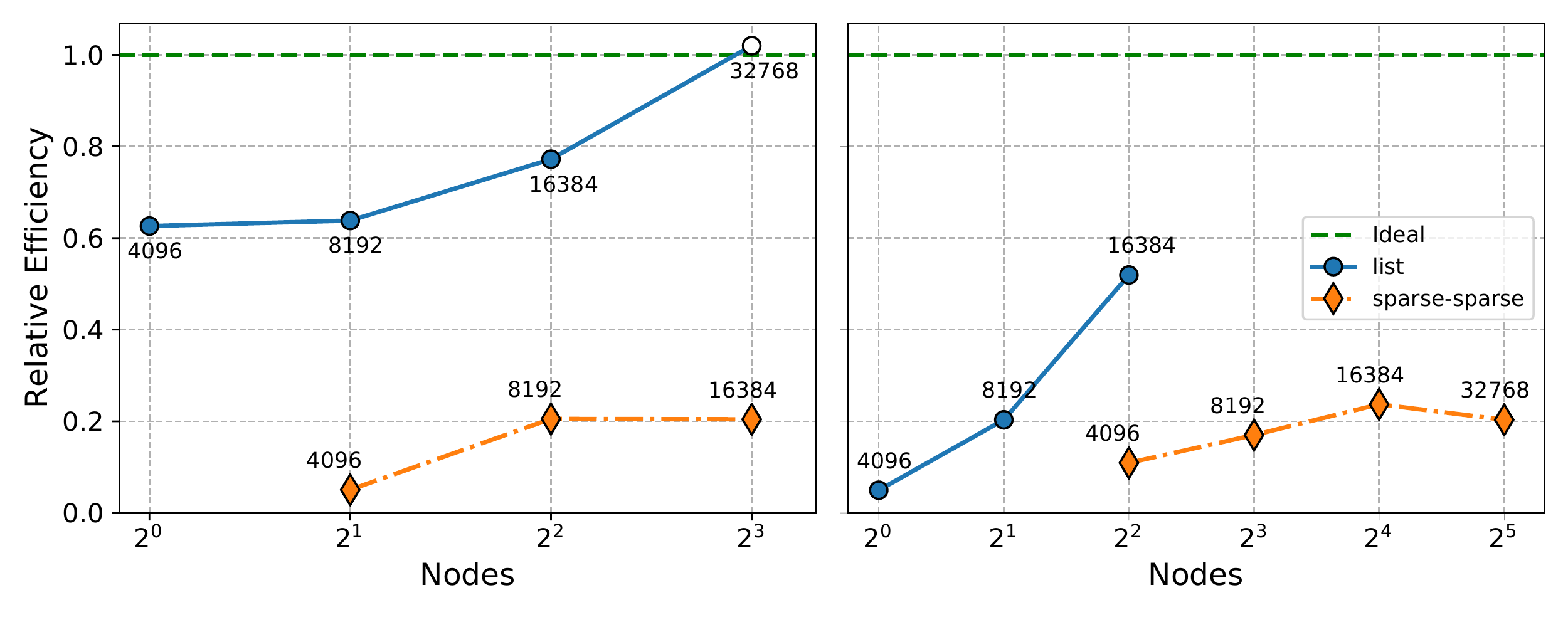}}
    \subfloat[Peak Relative Efficiency\label{fig:fermion_peak_eff}]{\includegraphics[width=0.85\columnwidth]{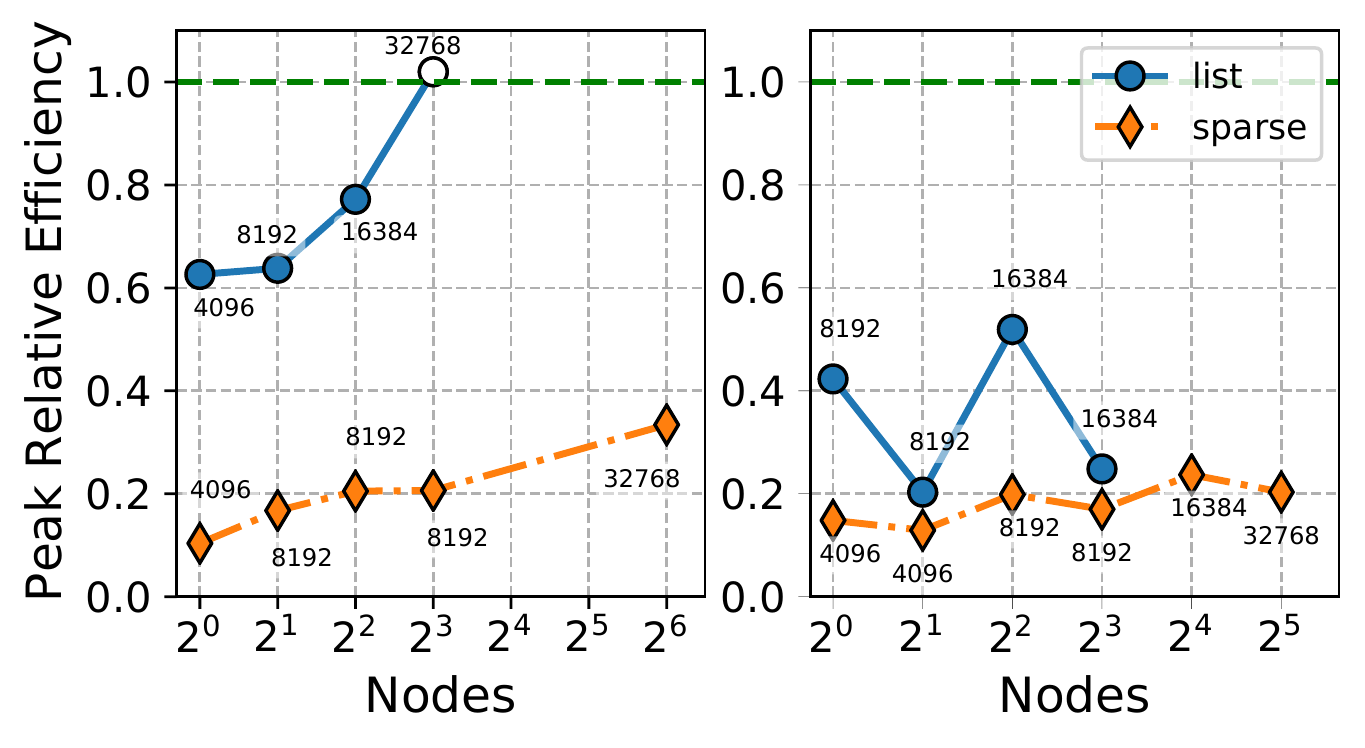}}
    \caption{Weak scaling measurements for electrons compared to single node ITensor for (a) fixed  $m/\text{node}$ and (b)  peak relative efficiency on Blue Waters (left plots of (a) and (b) ) and Stampede (right plots).  Relative efficiency is calculated by relative GFlops/s/node compared to single node ITensor at $m=16\thinspace384$ on Blue Waters and $m=8\thinspace192$ on Stampede2. Peak relative efficiency is taken as the highest relative efficiency measured at a given node count.  Bond dimension $m$ annotated at each point.    }
\end{figure*}
To determine time and flops/s, we again benchmark optimization of only one or two sites optimizations for each bond dimension. Finally, we construct the MPO with compression, where each order-4 tensor of $\tsr{H}$ is truncated via SVD to a $10^{-13}$ cutoff, resulting in an MPO with a bond dimension $k=26$. Similarly, the SVD truncation in the DMRG has an additional cutoff of $10^{-12}$ for $m<16384$ and $0$ otherwise.

\begin{figure}
    \centering
    \includegraphics[width=\columnwidth]{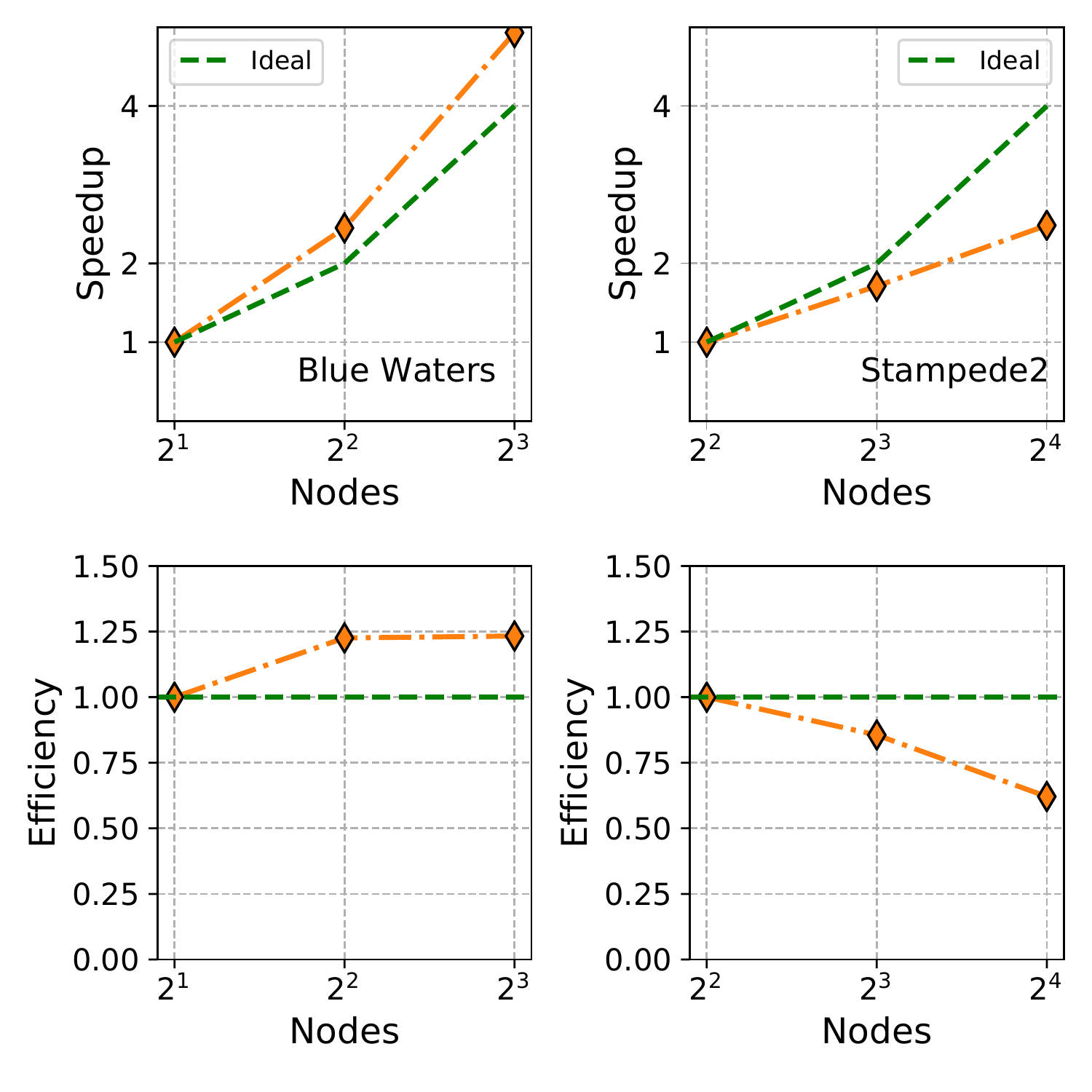}
    \caption{Strong scaling speedup (\textit{top}) and efficiency (\textit{bottom}) of Sparse-Sparse format for electrons at $m=8192$ on Blue Waters (\textit{left}) and Stampede2 (\textit{right}) }
    \label{fig:bw_ferm_strong}
\end{figure}

To begin, we benchmark sparse-sparse strong scaling performance on Blue Waters and Stampede2 in fig.~\ref{fig:bw_ferm_strong}. We see non-smooth performance due to communication contraction mapping infrastructure of Cyclops. Despite this, there is nearly ideal or better than ideal strong scaling speedup at $m=8192$ bond dimension. However, sparse format itself has a higher memory cost than the list format and thus on Stampede2 we find 4 nodes are the minimum rather than the 2 required on Blue Waters. 

Looking at DMRG weak scaling in fig.~\ref{fig:bw_fermion_weak}, we show that we gain efficiency only at the largest problem sizes\footnote{efficiency at small node counts may also be improved with larger problem sizes}. By comparing the relative time spent in fig.~\ref{fig:commCosts}(b), we find that the sparse-sparse algorithm is able to take more advantage of sparse MKL calls, while the list method is dominated by communication and CTF transposition costs. These sparse MKL calls grow from approximately 14\% time spent at $m=4\thinspace096$ to $52\%$ time spent at $m=32\thinspace768$ for the sparse-sparse algorithm on Stampede2. There is an additional dependence on architecture, further exemplified via the peak relative efficiency in fig.~\ref{fig:fermion_peak_eff}, as the sparse-sparse algorithm does not scale on Blue Waters, but is marginally better on Stampede2. 
The list algorithm is also more sensitive to overhead on each architecture: at the same node count Blue Waters has increased communication cost while Stampede2 has increased transposition costs.

\begin{figure}
    \centering
    \includegraphics[width=\columnwidth]{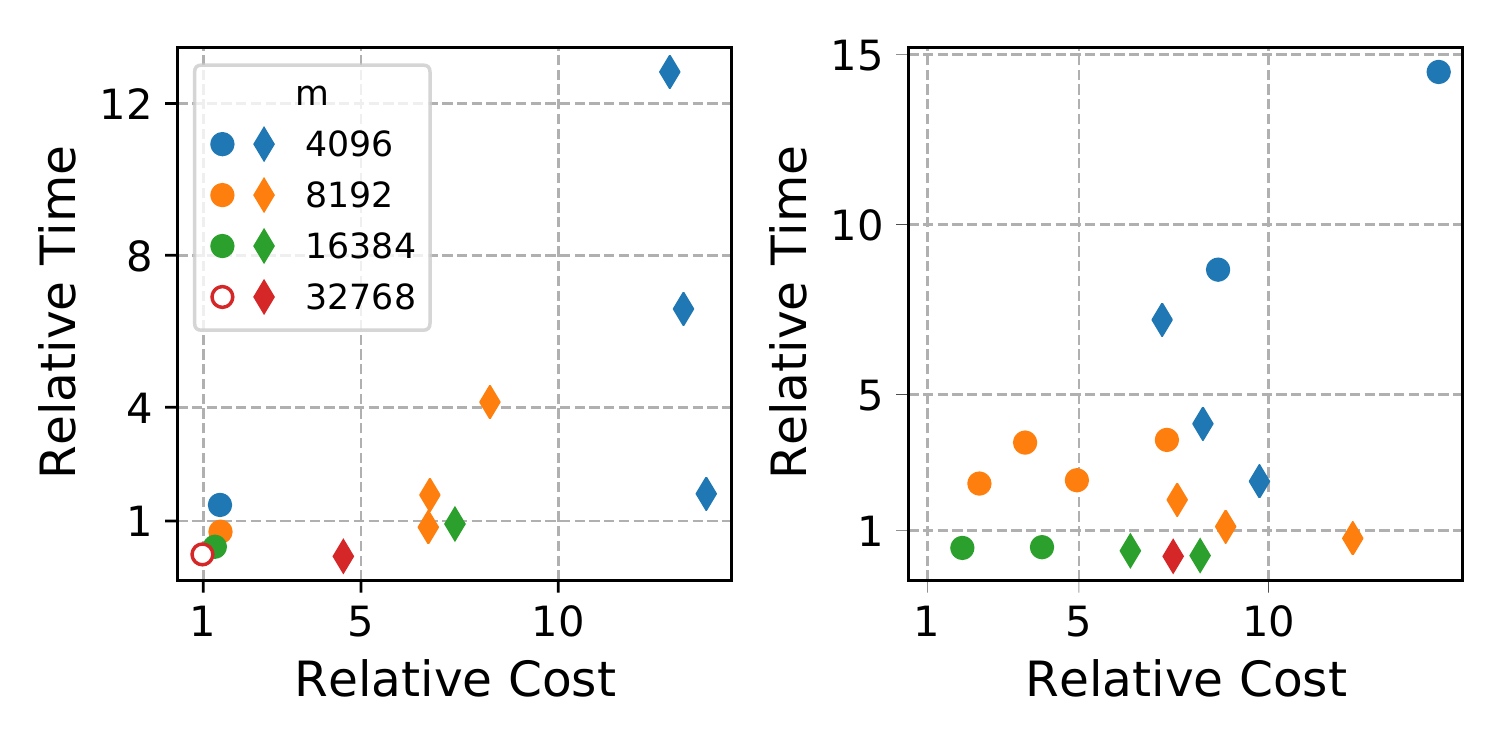}
    \caption{Electron system execution time and node hour cost relative to single node ITensor for list (circles) and diamonds (sparse-sparse) on Blue Waters \textit{left} and Stampede2 \textit{right} by varying hyperparameters ((node count and MPI processes/node).  Single node data was extrapolated past $m=8192$ for Stampede2 only. Open points are extrapolated from a half sweep.}
    \label{fig:fermion_time_cost}
\end{figure}

The discrepancy extends to relative time and cost of the two systems as well (see fig.~\ref{fig:fermion_time_cost}). For Blue Waters, the list algorithm is the only method that is efficient in both cost and time, where the largest problem has nearly a speedup of 8X and at nearly the same performance rate a serial node (0.98X). For the sparse-sparse algorithm, we see much more expensive time-to-performance, with $m=32\thinspace768$ receiving a 14X performance rate speedup at 4.5X the relative cost. 

On Stampede2, there is less of a drastic cost difference between algorithms. Using the list algorithm for $m=16\thinspace384$, there is a 2X performance rate improvement at a relative cost of 1.9X. The sparse algorithm exhibits a 3.9X speedup but at 8X the cost for our largest bond dimension. Time-to-performance can be lowered at low performance rate which leads to high costs.


\section{Conclusion}
We develop a distributed memory implementation of the DMRG algorithm and benchmark on two contrasting systems. Drastically different block structures in the different systems require different tensor block algorithms to be efficient at scale. 
For few, large blocks, we find excellent weak scaling efficiency for large problems and exhibit speedups that are nearly at serial cost and reduce time-to-solution by 5.9X. For many blocks, our two methods produce faster time-to-solution but at a worse performance rate than a serial node.
Despite these conflicting results, our implementation has an advantage over shared-memory models in that we can (weakly) scale to problems 64X greater in memory and 512X in complexity in real world applications. At up to 1.5X relative cost we were able to get up to 99X performance rate relative to an optimized single node code. This is crucial for solving the most complex problems with the high accuracy required to resolve conflicting results. 
These results should also extend to other DMRG-based algorithms and provide groundwork for future research on high performance implementations of tensor network methods. Particularly algorithms which do not need significant amounts of disk access are well suited for supercomputing applications at these large bond dimensions. 

\section{Acknowledgements}
We thank Xiongjie Yu who was the primary developer (under the guidance of BKC) of the serial version of the \TT~code on which our parallel version was built. We acknowledge both the use of ITensor to compare against our \TT~code as well as the use/modification of their AutoMPO and lattice code directly to facilitate comparison. This work used the Extreme Science and Engineering Discovery Environment (XSEDE), which is supported by National Science Foundation grant number ACI-1548562. We used XSEDE to employ Stampede2 at the Texas Advanced Computing Center (TACC) through allocation TG-CCR180006. This research is also part of the Blue Waters sustained-petascale computing project, which is supported by the National Science Foundation (awards OCI-0725070 and ACI-1238993) and the state of Illinois. Blue Waters is a joint effort of the University of Illinois at Urbana-Champaign and its National Center for Supercomputing Applications. BKC was supported by DOE de-sc0020165.
ES was supported by NSF grant no.\ 1839204.
% \clearpage 
\bibliographystyle{IEEEtran}
\bibliography{IEEEabrv,main.bib}
\end{document}